\documentclass[compsoc,conference,a4paper,10pt,times]{IEEEtran} 
\usepackage[utf8]{inputenc}
\usepackage[british]{babel}
\usepackage[utf8]{inputenc}
\usepackage[T1]{fontenc}
\usepackage{ae,aecompl}
\usepackage{amsfonts}
\usepackage{amsmath}
\usepackage{amsthm}
\usepackage{amssymb}
\usepackage{bm}
\usepackage{cite}
\usepackage{algorithmic}
\usepackage{algorithm}

\usepackage{anyfontsize}
\usepackage{times}
\usepackage{enumitem}

\usepackage{graphicx}
\usepackage{multirow}
\usepackage{bigstrut}

\usepackage{newfloat}
\DeclareFloatingEnvironment[placement={!ht},name=Algorithm]{mylist}
\usepackage[usestackEOL]{stackengine}
\strutlongstacks{T}

\usepackage{hhline}
\usepackage{lipsum}

\newcommand{\scell}[2]{\vtop{\hbox{\strut #1}\hbox{\strut #2}}}
\newcommand{\scellb}[2]{\vtop{\hbox{\strut \textbf{#1}}\hbox{\strut \textbf{#2}}}}

\let\svthefootnote\thefootnote
\newcommand\blankfootnote[1]{%
  \let\thefootnote\relax\footnotetext{#1}%
  \let\thefootnote\svthefootnote%
}
\let\svfootnote\footnote
\renewcommand\footnote[2][?]{%
  \if\relax#1\relax%
    \blankfootnote{#2}%
  \else%
    \if?#1\svfootnote{#2}\else\svfootnote[#1]{#2}\fi%
  \fi
}

\usepackage{etoolbox}
\BeforeBeginEnvironment{tabular}{\begin{center}\footnotesize}
\AfterEndEnvironment{tabular}{\end{center}}
\usepackage[colorlinks=true,urlcolor=black]{hyperref}

\def\BibTeX{{\rm B\kern-.05em{\sc i\kern-.025em b}\kern-.08em
    T\kern-.1667em\lower.7ex\hbox{E}\kern-.125emX}}

\title{Modelling direct messaging networks with multiple recipients for cyber deception}

\makeatletter
\newcommand{\linebreakand}{%
  \end{@IEEEauthorhalign}
  \hfill\mbox{}\par
  \mbox{}\hfill\begin{@IEEEauthorhalign}
}
\makeatother

\author{
    \IEEEauthorblockN{Kristen Moore\IEEEauthorrefmark{1}\IEEEauthorrefmark{2},
    Cody James Christopher\IEEEauthorrefmark{1}\IEEEauthorrefmark{2},
    David Liebowitz\IEEEauthorrefmark{3}\IEEEauthorrefmark{4},
    Surya Nepal\IEEEauthorrefmark{1}\IEEEauthorrefmark{2},
    Renee Selvey\IEEEauthorrefmark{1}\IEEEauthorrefmark{2}}
    \linebreakand
    \IEEEauthorblockA{
    \IEEEauthorrefmark{1}Data61, CSIRO\\
        Australia\\
        \{first.last\}@data61.csiro.au
    }
    \and
    \IEEEauthorblockA{
        \IEEEauthorrefmark{2}Cyber Security \\
        Cooperative Research\\
         Centre Australia\\
    }
    \and
    \IEEEauthorblockA{
    \IEEEauthorrefmark{3}Penten Pty Ltd\\
        Canberra, Australia\\
        \{first.last\}@penten.com
    }
    \and
    \IEEEauthorblockA{
        \IEEEauthorrefmark{4}UNSW\\
        Sydney, Australia\\
        \{first.last\}@unsw.edu.au
    }
}
\begin{document}
\def\tablename{\textsc{table}}
\maketitle

\begin{abstract} Cyber deception is the practice of deliberately introducing fake or misleading artefacts into cyber systems. It is emerging as a promising approach to defending networks and systems against attackers and data thieves. However, despite being relatively cheap to deploy~\cite{zhu2021survey}, the generation of realistic content at scale is very costly when it is hand-crafted. With recent improvements in Machine Learning, we now have the opportunity to bring scale and automation to the creation of realistic and enticing simulated content. In this work, we propose a framework to automate the generation of email and instant messaging-style group communications at scale. Such messaging platforms within organisations contain a lot of valuable information inside private communications and document attachments, making them an enticing target for an adversary. The presence of an active messaging platform also enhances the realism of a deceptive network simulation, contributing both traffic and message artefacts. We address two key aspects of simulating this type of system: modelling when and with whom participants communicate, and generating topical, multi-party text to populate simulated conversation threads. We present the LogNormMix-Net Temporal Point Process as an approach to the first of these, building upon the intensity-free modeling approach of Shchur et al.~\cite{shchur2019intensity} to create a generative model for unicast and multi-cast communications. We demonstrate the use of fine-tuned, pre-trained language models to generate convincing multi-party conversation threads. A live email server is simulated by uniting our LogNormMix-Net TPP (to generate the communication timestamp, sender and recipients) with the language model, which generates the contents of the multi-party email threads. We evaluate the generated content with respect to a number of realism-based properties, that encourage a model to learn to generate content that will engage the attention of an adversary to achieve a deception outcome. Our simulations run in real time, making them suitable for deployment in cyber deception as a honeypot in its own right, or as part of a larger deception environment.
\end{abstract}

\footnote[]{This work has been supported by the Cyber Security Research Centre Limited whose activities are partially funded by the Australian Government’s Cooperative Research Centres Programme.}
\footnote[]{Accompanying code for this work can be found at\\ https://bitbucket.csiro.au/projects/DECAAS/repos/emailgen/browse}
\section{Introduction}
\textbf{Cyber Deception.}
Deception is a deliberate attempt to manipulate the beliefs of others in order to influence their behaviour. The use of deception for cyber defence is known as cyber deception, and defined as \textit{"Planned actions taken to mislead and/or confuse attackers and to thereby cause them to take (or not take) specific actions that aid computer security defences"}~\cite{almeshekah2016cyber}. 


The predominant tool of cyber deception is the honeypot~\cite{spitzner2003-1-honeypots}, a fake system or digital media artefact which alerts defenders to breaches when attackers probe or interact with it. Honeypots can also provide information about the intent, and tactics, tools and procedures (TTPs) of attackers if they are {\em high interaction}~\cite{mokube2007honeypots}, with realistic detail and plausible behaviours. In particular, the content and behaviour of deceptive components, devices and data must all be similar enough to their real counterparts that they cannot be easily identified by attackers~\cite{zhu2021survey}.

One of the limitations of common honeypots is that they do not mimic the activity or interactions of their real counterparts, either because they do not have the capacity or are deployed in isolation. This makes them easier to identify as fakes. 
More sophisticated deceptions operate ensembles of simulations in a deceptive environment. Such an environment could include appliances and data that are likely to be attractive targets, and traffic-generating agents that interact with them. The realism of the data is important as part of the ``pocket litter" that support the deception\cite{stech2016integrating}. The interaction behaviour also generates traffic to maintain the illusion that the deception is a live appliance.

This approach is inspired by Cliff Stoll's famous use of deception to trap an intruder on the Lawrence Berkeley National Laboratory network in the late 1980s\cite{stoll1988stalking,stoll2005cuckoo}.
Stoll had been observing the intruder, but needed more engagement and connection time to allow international location tracing. Having determined their goals and interests, Stoll created a fictional computer network project that would attract attention. He generated a suite of documentation and correspondence between participant bureaucrats and military personnel. The ruse succeeded, and led to the arrest of the intruder. 

\textbf{Problem Statement.}
Despite the appetite for and demonstrated success of convincing simulations in cyber deception, their uptake has been inhibited by the fact that realistic generation at scale is very costly. Current deception technologies and tools exist on a continuum of sophistication, with simple but automated tools at one end and detailed, realistic and largely hand-crafted traps at the other. With recent improvements in Machine Learning we now have the opportunity to bring scale and automation to the creation of realistic and enticing simulated content.

This work investigates a model of what we will call a {\em direct messaging network} or DMN, encompassing both email and Instant Messaging (IM). IM apps now have more users than social network platforms\cite{Schaefer2016chat}, and though IM may have taken over email for social communications, within organisations email has evolved to become much more than just communications. Many email platforms are integrated with task management tools, and content management systems like Sharepoint, which encourage the sharing of important documents in the form of email attachments. This also results in users using email servers for personal information management, leaving information items as attachments in order to retrieve them later. For these reasons, simulated DMNs are a valuable addition to cyber deception systems. From the viewpoint of an adversary, they make an enticing target for discovering a company’s secrets, and accessing important documents that are shared between employees as attachments. 

This work is a step towards generating both realistic messages as a data resource and realistic multi-party interaction for common types of messaging systems like email and IM. Message transmission could be used to enhance the realism of a mail server honeypot by adding fake emails to its database and ensuring traffic to and from the appliance.  It could also be part of a larger deception in which some other appliance or data repository is the attractive target, and the messaging system provides a backdrop of plausible normal network activity.

IM and email are very similar from a modelling perspective. Communications in both originate with a message from a member of a network, addressed to a group of one or more members, often with a sequence of responses confined to that group. In this work, we model two separate but related aspects of a DMN:
\begin{enumerate}
    \item when and with whom participants communicate, and 
    \item the content of sequences of messages (threads or channels).
\end{enumerate}
We can simulate a DMN with the artefacts generated by these models, populating a messaging server with messages sent by agents on a network.

\textbf{Existing Approaches. }
Existing ML approaches~\cite{kimthesis} can model these two tasks, but there are limitations. Since DMNs model multi-cast communications, the natural setting to address the first of the DMN modelling tasks is to represent Direct Messages (DMs) as timestamped, directed hyperedges on the social network graph. Here the hyperedges are directed from one sender to one or more recipients. Accurate modeling of DMs therefore requires careful treatment of these hyperedges. Existing approaches to modeling DMs in the Temporal Point Process (TPP) literature~\cite{oPerry2013directed,kimthesis} model hyperedges as collections of independent edges. Formally, this is done by modeling edge weights for all pairs of nodes in the network, and then using binary classification independently on each edge incident to the sender node to decide whether or not to include the prospective recipient node in the recipient set. This dyadic network representation is a simplification with  analytical benefits, but higher-order dependencies seen in polyadic interactions like multi-cast DM communications can not be adequately expressed by these pairwise models\cite{chodrow2020configuration}. In particular, assembling the recipient set through independent selection can lead to over or underestimation of event rates for DMs with more than one recipient, and can also result in the construction of invalid participant sets that were not seen in the training data. To overcome this limitation, and increase the realism of generated deceptive communications, we propose the LogNormMix-Net Temporal Point Process, extending the intensity-free model developed by Shchur et al.~\cite{shchur2019intensity}. Our LogNormMix-Net learns edge weights for each hyperedge, as opposed to each pairwise edge as in prior works. 

\textbf{Proposed Solution.}
Our approach to generating the \emph{content} of communication threads leverages recent advances in deep learning based language models. There are a number of large pre-trained language generation models in the literature that are capable of generating coherent and grammatically correct textual content \cite{radford2019language,adiwardana2020humanlike,roller2020recipes,brown2020language, li2021pretrained}. Open AI's GPT-3 model~\cite{brown2020language}, which was released in 2020, has been found to be so good that without training, evaluators distinguished between GPT-3 and human-authored text at random chance level~\cite{clark2021all}. By fine-tuning such a pre-trained language model on an organisation's email or messaging dataset, communications can be generated that, in isolation, sound appropriate to the organisation. However this approach alone is not enough to simulate an office network where each individual will have consistent topics and themes to their communications, that are appropriate to their role in the organisation, and where communication threads are coherent and stay on topic. Such details are important to capture in order to fool increasingly sophisticated, AI-enabled adversaries. In order to achieve this, we propose a pipeline for fine-tuning large generative language models to generate coherent, topical, multi-party conversation threads. 

Our DMN model is then created by uniting the LogNormMix-Net TPP model with our multi-party conversation generation model. We validate our framework on real life data, and simulate an organisation’s email server, where each event generated with this system consists of a timestamp, sender, recipient set, email thread id, message type (new thread, reply, or forward), and body text. There is limited work in the literature on evaluating the realism of deceptive content and understanding how adversaries interact with deceptions. We propose a number of properties to evaluate the realism of the generated TPP content from our email server simulator, including preserving the distribution of inter-event times, the seasonality of communication interactions, and the proportion of events from/to each node/hyperedge in the network.   

\textbf{Contribution.}
Our main contribution is a framework for the end-to-end simulation of DMNs that can generate events in real time, making it suitable for deployment as a honeypot in its own right, or in larger deceptive environments. To the best of our knowledge, our proposed LogNormMix-Net TPP is the first neural TPP that models network communication events. It is also the first of any TPP approach (neural or parametric) where the recipients for multi-cast events are not independently sampled.
\section{Background}
There are two distinct components to our solution for simulating deceptive DMNs. For the first, we propose a neural TPP approach for modelling DMs within a social network. To develop our neural TPP, we draw upon foundational concepts from state-of-the-art methods found in the literature. This includes modeling communication network event streams with TPPs, and intensity-free approaches to learning TPP models. The second is to generate text content, for which we propose a pipeline to fine-tune large generative language models to produce coherent, topical, multi-party conversation threads. This section provides an overview of the background for both tasks.

\subsection{Content generation for cyber deception}
As outlined prior, content generation at scale for deception remains appealing yet under-investigated. With the increasing sophistication of attackers, there is a growing body of works introducing specialised generation models and attempting to measure their efficacy. As manual generation is expensive and does not scale, work has focused increasingly on automatic generation. Numerous algorithmic approaches exist that rely on existing files and artefacts~\cite{voris2012lost,bowen2009baiting, whitham2017automating,karuna-etal-2018-enhancing,karuna2018generating,li2021edge,timmer2021enticement,christopher2022dbschema}, however these suffer from a variety of drawbacks, including but not limited to decoherence issues and sensitive information leakage.
Recent works include models designed to generate coherent and cohesive long-form texts~\cite{cho2018towards}, and realistic software repositories~\cite{nguyen2021honeycode}.
Contemporary language models, such as the GPT family~\cite{radford2018improving, radford2019language, brown2020language}, enable the generation of novel and realistic text with appropriate fine-tuning and prompting, in addition to recent adaption for non-language domains~\cite{chen2020generative}. To the best of our knowledge, no work has been done in generating multi-party conversation threads.

\subsection{TPP theory and the LogNormMix model}\label{ifl-lognormmix}
Temporal point processes (TPP) are probabilistic models for event data with timestamps. Specifically, a TPP is a random process whose realization consists of the times, or equivalently the inter-event arrival times $\{\tau_i = t_i - t_{i-1}\}$, of isolated events. 
The simplest model of such event streams is the Poisson Process \cite{palm1943inten}, which assumes events occur independently of each other, with exponentially distributed inter-arrival times. Many extensions of the Poisson Process have been proposed to try to capture more complex relationships between event arrivals. Popular examples include the Cox Process~\cite{cox1955}, which consists of a doubly stochastic Poisson process, and the Hawkes Process~\cite{10.2307/2334319}, which consists of a self-exciting process that posits that past events temporarily increase the chance of future events.

A TPP is uniquely defined by its \textit{conditional intensity function} $\lambda^{*}(t):= \lambda(t\mid\mathcal{H}(t))$, which defines the rate of event occurrence given the event history, $\mathcal{H}(t)=\{t_i\}$. (Here the star is a shorthand notation that denotes the dependency on past events.) Formally, it’s the probability an event occurs in the interval $[t, t+dt)$ but not before $t$
\begin{equation*}
\lambda^{*}(t) = \dfrac{Pr\bigl(\text{event occurs in interval }[t, t+dt)\mid\mathcal{H}(t)\bigr)}{dt}
\end{equation*}

Conditional intensity functions provide a convenient way to specify point processes with a simple predefined behavior. The main challenge of intensity-based approaches lies in choosing a good form for the intensity function. Simpler intensity functions with closed-form log-likelihood tend to have limited expressiveness. More sophisticated intensity functions can better capture the dynamics of the system, but require approximation of the log-likelihood function using Monte Carlo or or numerical quadrature methods \cite{kimthesis,zuo2020transformer,zhang2020self}, which results in noisy approximation of gradients when training the model.\\

Shchur et al. \cite{shchur2019intensity} show that the drawbacks of the intensity-based TPP modeling approaches can be remedied by directly learning the conditional probability density function $p^{*}(\tau)$ of the time $\tau_i$ until the next event, instead of modeling the intensity function. The conditional intensity function can be expressed in terms of the probability density function $p^{*}(t) := p(t|\mathcal{H}(t))$ and its cumulative density function $F^{*}$ as follows
\begin{equation*}
\lambda^{*}(t) = \dfrac{p^{*}(t)}{1-F^{*}(t)}.
\end{equation*} 
This intensity-free learning approach for conditional density estimation\cite{shchur2019intensity} utilises normalizing flows to design flexible and efficient TPP models. The idea of normalizing flows is to define a flexible probability distribution by transforming a simple one. They propose using the LogNormMix model, a log-normal mixture model which estimates the density $p^{*}(\tau)$ using mixture distributions. Their LogNormMix model is implemented using the normalizing flows framework, but has the additional benefit of closed form sampling and moment computation.

\subsection{TPPs for modeling communication networks}
In addition modelling event times, TPPs can be extended to model additional event details such as types and/or locations. When there are two or more event types, each event in the TPP is described by a tuple $(\tau_i, m_i)$, where $\tau_i$ denotes the inter-arrival time and the event mark $m_i$ denotes the event type. TPPs have been used to model the timestamp and sender of social network communications, with the mark $m_i$ representing the sender \cite{fox2016enron, shchur2019intensity, zuo2020transformer}. DMN events are described by a timestamp, a sender, {\em and} one or more recipients, so in order to model DMN communications this standard TPP framework needs to be extended to include a recipient selection module. 

Accurate modeling of DMNs by adding recipient modeling to the classical TPP framework requires careful treatment of the hyperedges representing multi-cast events. Perry and Wolfe~\cite{oPerry2013directed} motivate their DMN model with a TPP intensity function that can model multi-cast event hyperedges. However for computational tractability reasons, they break multi-cast events into separate edges/events using duplication. Furthermore, they do not provide an algorithm for sampling multi-cast events using their approach.

To improve upon the duplication approach to dealing with multi-cast events, Kim et al.~\cite{kimthesis} proposed the hyperedge event model (HEM). HEM uses a dyadic modeling approach for recipient selection, placing recipient intensity models on the network edges. Hyperedges are then assembled as a function of these dyadic event participant functions. Namely, during event generation, the inclusion or exclusion of each edge is determined by drawing from an edge-specific Bernoulli distribution. In the typical \textit{binary relevance} formulation used by HEM, each single classifier operates independently, without any regard to other labels. This independent Bernoulli sampling is not ideal for modeling multi-cast events, as it is likely to result in the over or underestimation of true event rates, and even allows for recipient sets that have not been seen in the data to be sampled. To the best of our knowledge, the HEM of Kim~\cite{kimthesis} is the only work in the literature that includes a generative model for multi-cast events.

All existing works in the literature have used parametric TPPs to model network communications \cite{oPerry2013directed,kimthesis,fox2016enron,wu2019markovmodulated,ward2020network,price2019nonparametric}. Despite the elegance and simplicity of parametric TPP models, they assume that all samples obey a single parametric form, which is too restrictive and idealistic for real-world  data. As a result, neural network approaches have become popular in the TPP literature in recent years because of their universal approximation power, which enables them to (at least theoretically) generate a good approximation to the ground truth. To the best of our knowledge, no existing neural TPP work in the literature has been used to model network communications including recipients. There are no existing approaches that can model or generate communication network events involving both a sender and recipient(s), since they all lack a recipient selection module. In this work we extend the LogNormMix model of Shchur~\cite{shchur2019intensity} (described below) to model DM communications by adding a recipient module.

\subsection{Conversation generation using language models}
Controlled generation of textual content is a hot topic, and accordingly there is a multitude of work in this direction in the literature.  
One approach to generating conversation text is using open domain chatbots, such as Google's Meena~\cite{adiwardana2020humanlike}, or Facebook's BlenderBot~\cite{roller2020recipes}. They are typically trained on sequence-to-sequence (Seq2Seq) tasks using encoder-decoder ML architectures, making them suited to chit-chat or Q\&A tasks. Datasets in this Seq2Seq format tend to have short responses, and may not be suitable for more general applications.

Another option is to use one of the large pre-trained Transformers, of which OpenAI's Generative Pre-training (GPT) models~\cite{radford2018improving,radford2019language,brown2020language} are the most well known. These models can be used out of the box for language generation. Input text is given to the GPT model to prompt the generation, and the model will complete the text. Fine tuning can be used to bias generation to topics of interest. 

Plug and Play approaches~\cite{dathathri2019plug}\cite{duan-etal-2020-pre}\cite{mai-etal-2020-plug} aim to control the language generated from large, pre-trained language models. The Plug and Play Language Model (PPLM)~\cite{dathathri2019plug} employs small, custom attribute-based models to help control the language generated from general pre-trained language models. This results in cheap, but still powerful, conditional generative models. The control extends to the topic and sentiment of text.

\subsection{Simulating communication networks.}
DMNs can be modeled by uniting a TPP (to generate the communication timestamp, sender and recipients) with a language model (to generate the contents of the communication). There are a number of works in the literature that combine a TPP model with a topic model to model broadcast communications ~\cite{DBLP:conf/icml/HeRFGL15, bedathur2018discovering,topicDNHP2021}, however they do not include communication recipients in the modeling framework. To the best of our knowledge, the only work in the literature that models DMN events and content is that of Kim \cite{kimthesis}, who introduced the interaction-partitioned topic model (IPTM). IPTM integrates the HEM TPP model with the Latent Dirichlet allocation (LDA) topic model to identify the temporal and textual patterns in the data. This approach however has a number of limitations. The shortcomings of HEM's independent Bernoulli sampling of recipients is discussed above, and the topic model approach used by IPTM does not generate human readable conversation text. They instead use topic modelling to generate a set of word counts for each word in the vocabulary under consideration. This is a limitation that makes IPTM inappropriate for use in cyber deception, as an adversary that inspects a generated communication will immediately identify the deception.

We deliver a framework for the end-to-end simulation of DMN communications. Each event generated consists of a timestamp, sender, recipient set, conversation thread id, and the text of the communication. Unlike existing work, our TPP framework directly models hyperedge events, and our language model generates coherent conversation threads with topics personalised to the sender initiating the conversation.  

\subsection{Related Work}
\paragraph{TPP models of social networks}
There are TPP approaches in the literature designed to model public communications in social networks like Twitter or Sina Weibo (a Twitter-type platform in mainland China). Since they are modeling public, broadcast communications, recipients are not considered. A recent example is the Network Group Hawkes Process Model (NGH) of Xu et al. \cite{xu2020network}, which models network users by embedding network information into the Hawkes Process modeling framework. NGH captures the influence of a user's connected friends by adding an additional component to the intensity function, the network intensity, on top of the usual baseline and triggering kernel terms of the Hawkes Process. It is assumed that each user in the network comes from one of $G$ latent groups, according  to  their  dynamic  behavior  patterns. Users within a group are assumed to share the same set of group-level parameters, including the baseline intensity. Wu et al.~\cite{wu2020modeling} proposed the Graph Biased TPP (GBTPP) for modeling sequential event propagation on a network graph, such as retweeting by social network users, or news transmission between websites. 

\paragraph{Network TPP models that include recipients}
Network TPP approaches model communications between senders and recipients. An early example of this~\cite{oPerry2013directed} uses the multivariate Cox process to model email communications networks. The point processes are placed on the directed edges of the network to measure the rate of sending or receiving emails between pairs of email users. To extend this to cover multi-cast interactions between a sender and receiver set, they propose a multivariate intensity function. However, its log-partial-likelihood is complicated, so they instead use duplication to obtain pairwise interactions from the multi-cast events, and then evaluate the corresponding log-partial-likelihood of the pairwise intensity model as a way of performing approximate inference under the multi-cast model.

Fox et al. \cite{fox2016enron} presented a parametric Hawkes Process approach to model email network communications, where the parameters of the intensity function characterise important email communication behaviours such as the baseline sending rates, average reply rates, and average response times. In their approach, the point processes are placed on the nodes of the network, and the triggering kernel of the Hawkes Process term sums over only those messages received by the user, since the user cannot be influenced by emails they don't receive. Recipients aren't modeled in their Hawkes framework, and in the event generation algorithm for the model are sampled from a categorical distribution determined by counts from the training dataset. They propose to generate from their Hawks Process in layers, making it unsuitable for deployment in a system that requires real time event generation.

The Markov modulated Hawkes Process (MMHP) of Wu et al. \cite{wu2019markovmodulated} combines Markov modulation with the Hawkes process in an attempt to address the limitation of the Hawkes process in capturing “silent periods” and isolated events, whilst extending the flexibility of the Markov Modulated Poisson Process (MMPP) of Fischer and Meier-Hellstern \cite{fischer1993} to enable it to capture “bursty periods” of activity. 
Like in the multivariate Cox model approach of Perry and Wolfe \cite{oPerry2013directed}, pairwise intensity functions are placed on the directed edges of the networks. The model is motivated in terms of pairwise interactions in email networks, as well as fights between pairs of mice in a network of 12 mice living in a vivarium, and multi-cast events are not considered. 

Ward et al. \cite{ward2020network} propose the Cohort Markov-modulated Hawkes Process (C-MMHP) to improve upon the MMHP for modeling fights between pairs of mice. C-MMHP models the winner effect, where an animal that has experienced previous wins will continue to win future fights. This is captured in their model via a specific form to the coefficients of the triggering kernel in the Hawkes Process component. C-MMHP utilises degree correction in the pairwise baseline intensity to allow the model to better capture individual level heterogeneity. This work only models pairwise events, and multi-cast events are not considered. 

\section{Key Insights: multi-cast event modeling}\label{section:motivation}
Temporal Point Processes (TPPs) have been very popular for modeling various aspects of public, broadcast interactions on social networks like Twitter, Reddit and Yelp~\cite{shchur2019intensity,zuo2020transformer,topicDNHP2021}. In this work, we use TPPs to model and simulate private communications within organisations, which take place on email servers, messaging apps, and communications platforms like Teams and Slack. The key aspect differentiating these two types of communication is the recipient set. Recipients are not considered when modeling broadcast, newsfeed-style messages, as everyone following the person posting the message is considered a recipient by default. On the other hand, DMs are multi-cast events, where the sender selects a set of recipients from their social network every time a message is sent. Accurate modeling of recipient selection in multi-cast communications is therefore critical for the simulation of realistic DMN communications for cyber deception.

Since DM platforms allow for multi-cast communications, the natural setting for DMs is to represent them as timestamped, directed hyperedges on the DMN graph. As discussed, the weakness of existing DMN TPP models~\cite{oPerry2013directed,kimthesis} is that they model hyperedges as collections of independent edges. To better understand the implications of this modelling approach, we consider a toy example of a fictitious organisation E Corp. Logically, the CEO of E Corp sends many messages to the Chairman of the Board and also to their team of C-level executives (including the chief operating officer, chief financial officer, and chief marketing officer), but they do not typically send the same messages to both the Chairman and the C-level executives. In this section we motivate why such higher-order dependencies can not be adequately expressed by existing pairwise models, and how our hyperedge modelling approach can address this limitation.

\begin{figure}
\includegraphics[width=\linewidth]{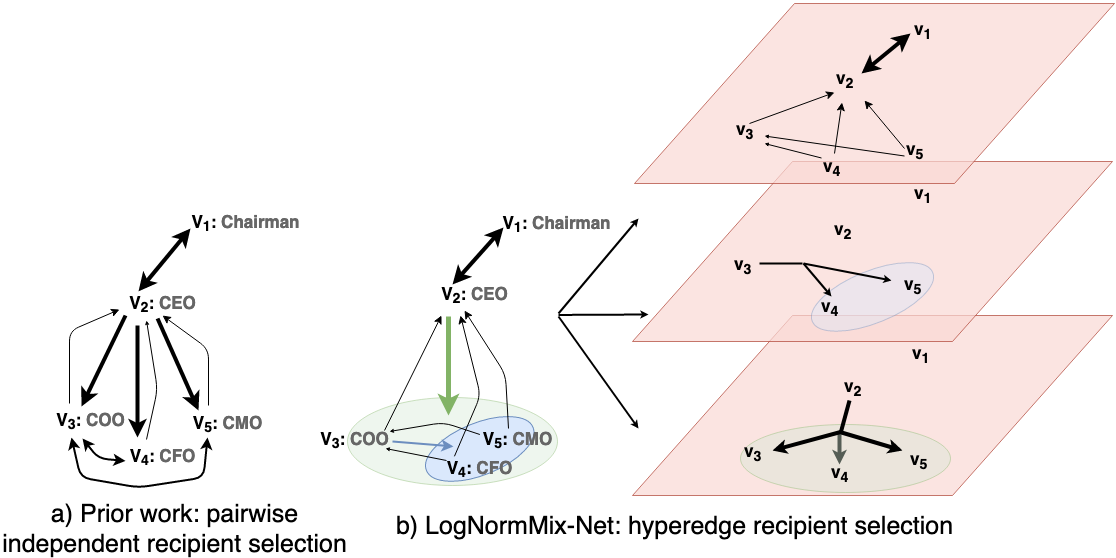}
\caption{Graphical representation of two different approaches to modelling DMs at the fictitious company E Corp. Figure (a) shows the binary classification approach to recipient modeling is able to capture the pairwise probability of a communication being sent between 2 nodes. Figure (b) shows the multi-class classification approach is able to capture the probability of a communication between a sender and an entire recipient set.}\label{toy-problem}
\end{figure}

Figure~\ref{toy-problem}(a) depicts the edge weights learned via the pairwise approach of existing works, which independently assemble recipient sets using binary classification. Figure~\ref{toy-problem}(a) shows that this approach can capture the fact that the CEO communicates with the Chairman and each of the C-level executives roughly the same amount, illustrated by the fact that the outgoing edges from \textbf{v}$_2$ are the same width. It is however unable to capture the fact that the CEO and the Chairman of the Board interact solely via unicast communications, whereas communications sent by the CEO to the C-level executives are all multi-cast. This communication behaviour is clearly evident in Figure \ref{toy-problem}(b), which illustrates the hyperedge modelling approach we propose in our LogNormMix-Net model. To address the limitation of previous works, our model learns edge weights for each hyperedge. This is achieved by modeling recipient selection as a multi-class classification problem, where the classes are composed of the distinct recipient sets seen in training.

To understand how recipients are chosen during event simulation in both approaches, we consider the case where the CEO at node \textbf{v}$_2$ is sending an email. Under the binary classification approach of Figure \ref{toy-problem}(a), the model will estimate a pairwise probability that each of \textbf{v}$_1$, \textbf{v}$_3$, \textbf{v}$_4$ and \textbf{v}$_5$, will be included as a recipient. An independent Bernoulli sample will then be drawn from each of these 4 pairwise probabilities to decide whether or not the respective person/node is included as a recipient. The problem with this is that each sample is unaware of the outcome of the other 3 samples, whereas Figure \ref{toy-problem}(b) shows that if the Chairman at node \textbf{v}$_1$ is selected as a recipient, then none of the C-level executives should be chosen. On the other hand, if one of the C-level executives are chosen, then they all should be chosen, since the CEO's only communications to nodes \textbf{v}$_3$, \textbf{v}$_4$ and \textbf{v}$_5$ are multi-cast messages sent to all of them together. This example highlights how this multi-label binary classification approach can result in over- or under-estimation of true event rates for hyperedges involving more than one recipient, and can even allow for invalid recipient sets to be selected. Since such higher-order dependencies can not be expressed by pairwise intensity models, this motivates our alternative approach to modelling hyperedge events.

Our LogNormMix-Net approach frames recipient selection as multi-class classification problem, conditioned on the sender of the communication. This enables the model to learn the probability distribution of the recipient sets, given the sender - which is akin to learning hyperedge weights for directed communications between participant sets. A further benefit of this approach is that only those recipient sets seen in training are available for the model to select, thereby preventing invalid recipient sets from being constructed.
\section{LogNormMix-net for modeling DMNs}\label{TPP-model}

In this section, we present the technical details of how we extend upon the LogNormMix model~\cite{shchur2019intensity} to create our network communication TPP model to generate deceptive communications at scale for cyber defence. 

\subsection{Why choose LogNormMix?}
Before discussing our model architecture, it is pertinent to justify our choice to build upon the LogNormMix model\cite{shchur2019intensity}. There are 3 properties that are desirable when designing any TPP model~\cite{shchur2021neural}:
\begin{enumerate}[label=(\roman*)]
    \item Flexibility: a TPP should have the ability to approximate any probability density on $\mathbb{R}$ arbitrarily well, including multi-modal ones.
    \item Closed form likelihood: in cases where a closed form of the likelihood function cannot be computed, approximation via Monte Carlo or numerical quadrature must be performed, which is slower and less accurate.
    \item Closed form sampling: closed form sampling is especially important for this work. In particular, it is ideal to be able to draw samples analytically via inversion sampling\cite{rasmussen2011temporal}. In cases where a closed-form expression of the inter-event time distribution is not available, sampling is difficult to do. A popular approach in this case is to use the thinning algorithm \cite{lewis1978thin}, which involves generating a sequence of events from a homogeneous Poisson process with intensity given by the upper bound of the intensity function you wish to sample from, then rejecting some of the events so that the sequence follows the desired point process. If the upper bound on the intensity function is not tight, a large fraction of samples will be rejected. This approach is less accurate and slower than inversion sampling, and is not suitable for parallel hardware like GPUs.
\end{enumerate}

There are two existing approaches that satisfy all three of these properties. The first is specifying the PDF $p^{*}(t)$ with a mixture distribution~\cite{shchur2019intensity}, and the second is with invertible splines~\cite{shchur2020fast}. In this work we choose to build upon the mixture distribution approach, as it is naturally extendable to network communication modeling whilst preserving all of these properties, whereas the existing invertible splines approach does not model marked events.

\subsection{Developing a network event TPP model}
To model DM communication events in a social network of size $n$, we represent an event by the tuple $(\tau_i, s_i, r_i)$, where $\tau_i$ denotes the inter-arrival time, $s_i$ denotes the sender (represented by an integer between $1$ and $n$), and $r_i$ denotes the recipient list, represented by an $n$-dimensional multi-hot vector with unit entries denoting the recipients for event $i$. 

\begin{figure}[htb]
\includegraphics[width=8cm]{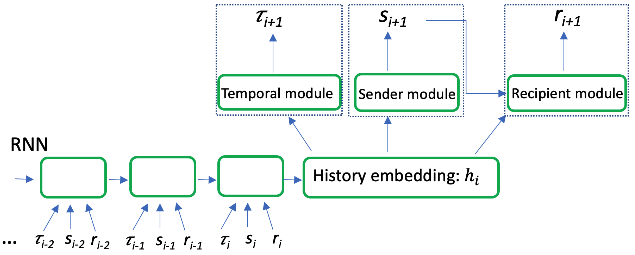}
\caption{High level overview of the LogNormMix-Net architecture for event simulation.}\label{basic-architecture}
\end{figure}

We frame network event modeling as a multitask machine learning problem with 3 tasks: inter-arrival time prediction, sender prediction, and recipient prediction. Each of these tasks has its own prediction module, and associated negative log likelihood loss function:  
\begin{equation}\mathcal{L}_{total} = \sum_{i}l_{\tau_i} + l_{s_i} + l_{r_i}\end{equation} 
The event history, $\mathcal{H}=\{\tau_i, s_i, r_i\}$ is embedded with an RNN into a fixed-dimensional vector $h_i\in\mathcal{R}^H$, and then input into each of the 3 modules, as shown in Figure \ref{basic-architecture}.

As in the original LogNormMix implementation, inter-event times are conditionally independent of the event participants, given the history. When extending the LogNormMix to the network communication setting, we frame event participant selection as a sender-driven process, i.e. the sender is chosen first using the sender module. Then given the sender, recipient selection is formulated as:
\begin{enumerate}
    \item a multi-class classification problem
    \item conditioned on the sender
\end{enumerate}
Detailed motivation for this formulation for the recipient selection module is given in Section \ref{section:motivation}. 

\subsection{LogNormMix-Net model architecture}
In this section we provide the technical details of each of the modules in the LogNormMix-Net. 

\textbf{Temporal module}
As in the LogNormMix model, the inter-arrival time probability distribution $p^{*}(\tau)$ is modeled by a log-normal mixture model 
\begin{equation}
\label{lognormal-mixture}
    p(\tau\vert\omega, \mu, \sigma) = \sum_{k=1}^K \omega_k \dfrac{1}{\tau \sigma_k \sqrt{2\pi}} \exp{-\dfrac{(\log \tau - \mu_{k})^2}{2\sigma_k^2}}
\end{equation}
where $K$ is the number of mixture components, $\omega$ denotes the mixtures weights, and $\mu$ and $\sigma$ are the mixture means and standard deviations, respectively. In order to help the model accurately learn the seasonality of DM sending in the training data, metadata $m_i$ is concatenated with the history embedding vector $h_i$ to serve as additional input to the model. The choice of metadata is dataset specific - for instance, the distribution of sending times of emails on a corporate network depends on the day of the week and whether or not the current time is within office hours.  

\textbf{Sender module}
The sender module consists of a feedforward network plus softmax. The history embedding vector $h_i$ is input into the sender module, and the resulting output logits define a categorical distribution across all the nodes in the network. For prediction tasks, the next sender $s_{i+1}$ is predicted to be the class with the highest probability. In our experiments the feedforward network consists of two fully connected layers separated by a tanh activation layer.

\textbf{Recipient module}
The recipient selection module is an additional module that we have added to the original LogNormMix model. We formulate recipient selection as a multi-class classification problem by creating an ID for each of the recipient set combinations seen in the training set. Then similar to the sender module, a fully connected layer plus softmax is used to learn a categorical distribution across all the recipient IDs in the dataset. In order to condition the recipient selection probability distribution on the sender, we concatenate the history context embedding with the sender embedding (ie. the embedding of the ground truth sender), and pass that as input into the recipient selection module. As with the sender module, for prediction tasks, the next recipient $r_{i+1}$ is predicted to be the class with the highest probability.

\subsection{Event generation with LogNormMix-Net}
The LogNormMix-Net architecture makes it simple to sample event streams. The next inter-event time $\tau$ is sampled from the lognormal mixture in equation \ref{lognormal-mixture} using the standard mixture model sampling approach:
\begin{enumerate}
    \item Sample the mixture component:\\ 
    $ \bf{z}\sim Categorical(\bf{\omega})$\\
    where $\bf{z}$ is a one-hot vector of size $K$.
    \item Sample from the unit Normal distribution: \\
            $\varepsilon\sim Normal(0,1)$.
    \item Transform the unit normal sample $\varepsilon$ to the lognormal distribution for mixture component $\bf{z}$: \\
            $\tau = exp(\bf{s^{T}z}\cdot\varepsilon + \bf{\mu^{T}z})$.
\end{enumerate}

The sender and recipient(s) of the next email is drawn from their respective categorical distribution, with the sampled sender being fed as input into the recipient selection module. A detailed diagram of the use of our LogNormMix-Net model for event sampling is shown in Figure \ref{conditionalarch} in Appendix \ref{lognormmix-arch}. 

\section{Multi-party conversation generation}\label{text-gen}

The other key task of our DMN simulation solution is to generate multi-party conversation threads. To do this we leverage generative pre-trained language models, where the input prompt text is designed to personalise and steer the topic of generated emails. We validate our approach in Section \ref{experiments} using GPT-2 , though our approach does not depend on any particular model. Due to the private nature of their contents, limited DM datasets are available, so we validate our approach by simulating email threads using the Enron email corpus. Our implementation requires fine-tuning the pre-trained generative language model on the Enron email dataset, however this dataset need not be labelled.

\subsection{Email content generation}
A difference between email and other DMN platforms like Whatsapp, Teams or Slack, is that emails include a subject line. When simulating an email server, our best results were obtained when an additional, separate model for email subject generation was used. In particular, we train 2 generative language models:
\begin{enumerate}
    \item \textit{Email Subject generator model:} In cases where the sender is starting a new email thread, this language model is used to generate the subject line. For reply or forwarding emails, the subject is inherited from the existing email thread, and is therefore not generated.
    \item \textit{Email body generator model:} This language model generates message for the body text for each email.
\end{enumerate}

The process to train and utilize the subject and body generation models is outlined below. 

\noindent\textbf{1) Email Subject generator model:}
\begin{itemize}
    \item[--] Training: the model is trained by fine-tuning a pre-trained language model on the email subject lines from the Enron corpus. 
    A set of topic/keywords are also extracted for each user, that represent topics of interest to them.
    \item[--] Generation:  In order to personalise email subject generation, the topic of conversation threads is controlled by using a topic word that is sampled from the sender's keywords as the input prompt text to the subject generation model.
\end{itemize}

\noindent\textbf{2) Email body generator model:}
\begin{itemize}
    \item[--] Training: the body generation model is trained by fine-tuning a pre-trained language model on the email bodies from the Enron corpus.
    \item[--] Generation: for a new thread, we use the email subject (which we generate before the body) as the input prompt text for the email body generation model. In the case of reply/fwd emails, the subject and existing email thread text is used as input prompt text.
\end{itemize}

\subsection{Assembling email threads}
\label{subsec::emailgen}
\textbf{Sampling the communication type}
In order to develop realistic email threads, 3 types of communication are modeled: 
\begin{enumerate}[label=(\roman*)]
    \item "new thread": starting a new conversation thread,
    \item "reply": replying in an existing conversation thread,
    \item "fwd": forwarding the existing communication to a new recipient. 
\end{enumerate}    
For each sender, we try to ensure the proportions of each communication type (new-thread, reply, fwd) in the rolling simulated content resembles the proportions of that sender from the training data. The pseudocode for email type selection is provided in Appendix \ref{FirstAppendix}.

\textbf{Email thread generation}\label{emailalgo} 
The detailed steps required to generate an event, consisting of (email ID, communication type, thread ID, email subject, email body), assuming the email sender and recipients are known, are detailed in Appendix \ref{email thread algo}. emails with the same thread ID constitute a communication thread, with the message order within the tread implied by the ordering of the email ID. We make use of canned text where appropriate, by sampling from canned lists for greetings, salutations and for the message text for forwarded emails. Figure \ref{flowchart} gives a high level overview of the main components of the framework.\\

\subsection{Additional implementation options}

\textbf{Increasing the realism of generated content} 
There are various strategies that one may take when preparing the training dataset. To increase the realism of generated text to be used in cyber deception, it would be ideal to fine-tune the pre-trained generative language model on real, but non-sensitive communications from the organisation being simulated. If real data is not available, it is possible to use reg-ex and NLP approaches to create a training data pre-processing script that can replace Enron-specific terms like names of entities and locations, etc.,  with those relevant to the organisation wishing to deploy the generated content. 

To try to improve the cohesion of communication threads, a training dataset was created that captures the existing conversation thread for each training sample, rather than treating email bodies independently of any existing communications.

\textbf{Increasing the enticement of generated content}
Another option is to employ "Plug and Play" language models. We investigated the Plug and Play Language Model (PPLM)~\cite{dathathri2019plug} as an approach to give better control over the topic of the generated emails, with a view to increasing their enticement. This would be especially useful in cyber deception when a defender can anticipate a key piece of information that an adversary would be interested in, or some other aspect of their goals and perceptions~\cite{timmer2021enticement}. PPLM enables that information to be leveraged, by adding relevant words to the PPLM model's bag of words (BOW), which then encourages the model to use those words during generation. An intruder navigating the file system, searching for files worth stealing, would then surface the simulated decoys that had the keywords in them~\cite{salem2011decoy}.

\subsection{End-to-end simulation of social network communications}
\begin{figure}[htb]
\centering
\includegraphics[width=7.5cm]{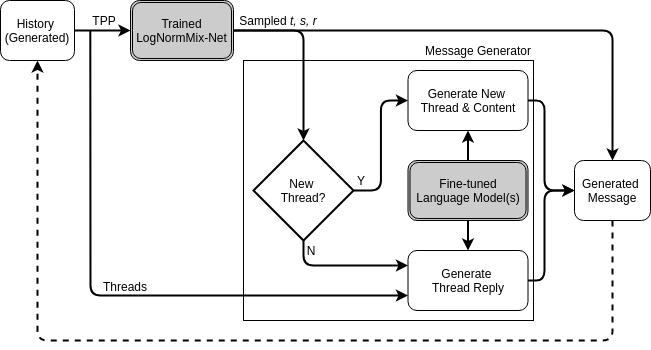}
\caption{Overview of our email generation framework.}
\label{flowchart}
\end{figure}

Our framework for the end-to-end, real-time simulation of DMNs now follows simply by uniting our LogNormMix-Net TPP from section \ref{TPP-model} with the language models from section \ref{text-gen}. Figure \ref{flowchart} gives a high level overview of the approach.

To summarise the key steps required to generate an event, consisting of (timestamp, sender, recipient set, email ID, communication type, thread ID, email subject, email body); \emph{(a)} Generate the timestamp, sender and recipients using the LogNormMix-Net TPP model, and \emph{(b)} Generate the email ID, communication type, thread ID, email subject and email body as detailed in~\S\ref{subsec::emailgen}.
\section{Experiments}\label{experiments}
There is limited work in the literature on understanding how adversaries perceive and interact with deceptions, and how the realism and enticement of deceptive content can be quantified\cite{karuna-etal-2018-enhancing,karuna2018generating,timmer2021enticement}. We posit that an adversary who is being careful not to betray their presence on a system does not have the luxury of being able to thoroughly inspect every aspect of a network environment for signs of imperfection indicating deceptive content. For this reason, generated deceptive content need not be perfect, but should stand up to moderate scrutiny. For our messaging application, if an attacker has compromised a user account and is searching the email server for keywords, we want the preview snippet of text returned by the search to be plausibly realistic. This means that \textit{local} coherence and cohesion of generated textual content is a minimum requirement. Additionally, we ask that the simulated interaction events from our TPP model appear to be generated from the same distribution as the training data, so that interactions patterns and temporal dynamics are preserved.

Our experimental aims in this section are to validate:
\begin{enumerate}
    \item our choice to use the multi-class classification recipient selection approach, over multi-label binary classification,
    \item that our end-to-end email simulation framework can generate convincing content, appropriate for application in cyber deception. 
\end{enumerate}

We present a combination of statistical goodness of fit tests from the literature, together with realism tests specific to our application of generating realistic deceptive content. 

We train TPP models on 2 real-world email datasets. Each dataset consists of multiple sequences of events, and we use 60\% of the sequences for train, 20\% for dev and 20\% for test. For the generation of email text, we train our model on the Enron email corpus.

\subsection{Experimental Setup and Metrics}
\subsubsection{LogNormMix-Net evaluation}
When training our TPP models, we minimize the combined negative log-likelihood (NLL) loss of the inter-event times, the next-sender prediction and the next-recipient prediction in the training set. We then evaluate our trained TPP model on 2 tasks: event prediction and generation.\\
\noindent\textbf{TPP Event Prediction} Event prediction is a classic goodness of fit test that is popular in the neural TPP literature. Though this task is not directly related to our cyber deception application, we include it in order to compare the ability of the two different recipient classification approaches to learn the relationship between senders and the recipient groups they communicate with. In particular, the recipient prediction task asks the model which recipient set is likely to be chosen next, given the sender of the next communication, and the history of all previous communications.

As is customary in the neural TPP literature, we report the NLL loss for event time prediction, sender prediction and recipient prediction on the test set. In addition to this, for the time prediction task we report the RMSE and MAE, and for sender and recipient prediction we report the top-1 and top-3 accuracy. \\
\noindent\textbf{TPP Event generation}
For this task, we perform 100 simulations for each dataset. We then compare the generated event sequences with the ground truth training data. To evaluate the temporal realism, we calculate the Q-Q plot and Earth Mover's Distance (EMD) of the inter-arrival time distribution. In addition to this, we perform seasonality checks by plotting histograms and computing the EMD of the proportion of emails sent in each of the 24 hours of the day, and 7 days of week. 

To evaluate the realism of the participant selection, we compute the EMD of the proportion of email sent by each sender and the proportion of emails received by each recipient group. 

\subsubsection{Email content evaluation}
The email subject and body generation models were trained using the Huggingface implementation of GPT-2\footnote{\url{https://huggingface.co/transformers/model_doc/gpt2.html}}. 
We then created a corpus of email subjects. For each person in the network, we took the 10 most frequently used words from their keyword list, and then used each one as a prompt to generate roughly 45 subjects. This resulted in a set of at least 450 subjects generated per user. This is used to investigate whether using keywords as prompts for the subject generation can differentiate sender’s emails from one another, and give an indication of their role in the network. A corpus of email threads was then generated following the Algorithm in~\S\ref{subsec::emailgen}.

Metric-based approaches for evaluating the realism of text in the literature aim to evaluate coherence and cohesion. Ideally, for this work we would like to have such a metric that can quantify coherence and cohesion of a sequence of messages in a communication thread. We are not aware of any such metric in the literature, and so have taken a similar approach to that in Karuna et al. \cite{karuna-etal-2018-enhancing}, who propose a novel method to enhance cohesion and coherence of fake text used in cyber deception, in order to improve its believability. In \cite{karuna-etal-2018-enhancing}, they measure the semantic coherence of a document by averaging the similarity between the adjacent sentences or paragraphs. Since coherence is a necessary requirement for realism in deceptive content, in this work we have reported similarity scores between 2 consecutive emails within a thread. 

To calculate similarity between 2 emails, an email embedding vector was created for each email using Google's Universal Sentence Encoder\cite{cer2018use}, then cosine-similarity was used to measure the semantic similarity between the email embedding vectors of the 2 consecutive emails in a thread. Since cosine-similarity is used, the maximum possible similarity score is 1 and the minimum is -1, though negative scores are rarely observed. 

Referential cohesion is used to measure the overlap of ideas across adjacent bodies of text. In \cite{karuna-etal-2018-enhancing}, cohesion is calculated as the linguistic overlap of words across adjacent paragraphs. Following this, we calculate the cohesion between an email and its reply as the number of overlapping lemma types that occur in both emails, which we compute using spaCy's \textit{en\_core\_web\_lg} model.

\subsection{Datasets}
For each of the datasets we extract the timestamp, sender, recipients, and metadata for each event, where the metadata is a categorical input we compute from the date-times, that denotes whether the event happened during typical office hours, shoulder period (weekday morning and evenings ouside of regular office hours), or during non-working hours (weekends plus weekday nights).

The \textit{EU Email dataset}\cite{paranjape2017motifs} consists of the email network from a large European research institution, divided into 4 departments. To create a smaller network from the department 4 dataset, we selected the 52 members who sent at least 100 emails over the duration of the dataset, and then created a closed network of the emails sent between them. This comprised 21,260 emails sent between the 52 employees over a period of approximately 75 weeks beginning in October 2003. From this we discarded 907 emails that were sent to recipient groups that received fewer than 10 emails in the entire dataset. 
This resulted in a training set of 20,353 emails sent between 52 employees and 123 different recipient sets.  Of these emails, 68.4\% were multi-cast, ie. having 2 or more recipients.  From the 75 weeks of data, 6 weeks of low activity (less than 80 emails) were discarded, leaving 69 weeks of data comprising 20,098 emails.\

The \textit{Enron Email dataset}.  
The Enron dataset was originally made public, and posted to the web, by the Federal Energy Regulatory Commission during its investigation into the Enron corporation. The CMU version of the dataset was used in this work, which does not include attachments, and has had some messages deleted "as part of a redaction effort due to requests from affected employees"\footnote{\url{https://www.cs.cmu.edu/~./enron/}}. For our TPP work, we consider the sender, recipient, and timestamp of each message in a closed version of the Enron email network containing messages sent between 148 users. Once duplicates and messages individuals sent to themselves are removed, the corpus is reduced to 30,021 unique messages. From this, a smaller network was created by selecting the 54 users who each sent at least 100 emails, and then forming a closed network from those between 164 different recipient groups. This resulted in a dataset of 13,726 unique messages, spanning 144 weeks. 

\textbf{Ethics}
Due to the sensitivity of the Enron emails, we obtained an ethical clearance for use of the Enron dataset in our work. Since the other 2 datasets consist only of timestamps and anonymous participant ID numbers, we consider them to be fully anonymised.

\subsection{Baseline models/comparisons}
\subsubsection{TPP event prediction}
\textbf{LogNormMix}
We report the performance of the original LogNormMix model for inter-arrival times and sender prediction on our DM event datasets. This allows us to set a baseline for temporal and sender prediction performance in order to demonstrate that adding the recipient prediction task doesn't degrade performance on these original tasks too much.\\
\textbf{Transformer Hawkes}
The Transformer Hawkes Process (THP)~\cite{zuo2020transformer} is another SOTA neural TPP model. We report the RMSE and sender accuracy of this model to serve as a comparison with the baseline LogNormMix.\\
\textbf{LogNormMix with binary classification: LNM-BC}
In order to compare the performance of the multi-class and multi-label binary classification approaches for the recipient module, we include LNM-BC, which is the LogNormMix with a recipient module that consists of the typical multi-label binary classification approach (shown in red in Figure \ref{conditionalarch}).

\subsubsection{TPP event generation} 
We evaluate the generated content from our LongNormMix-Net against 2 other models. Firstly, the LogNormMix with binary classification (LNM-BC) described in the paragraph above. The second comparison model is a parametric TPP: the Hawkes TPP model of Fox et al. \cite{fox2016enron}. This parametric TPP approach does not model recipients, and therefore does not constitute a network TPP in the sense that we consider in this paper. However, their paper does include an algorithm to sample uni-cast communications. In this work, we (trivially) extend their sampling algorithm to enable it to sample multi-cast communications.

\subsubsection{Conversation thread generation}
Since there are no existing approaches in the literature to evaluate the realism of communication threads (to the best of our knowledge), we evaluate our generated text by simply comparing the coherence and cohesion of our generated content against the scores for the ground truth email threads from the Enron dataset (that we fine-tuned our GPT-2 model on). To create a test set to use for this comparison, we selected 10 emails threads from the Enron emails corpus at random, comprising a total of 39 emails.

\section{Results}\label{results-section}

\subsection{LogNormMix-Net}
In section \ref{goodness-of-fit} we address our first experimental aim, using the event prediction task to validate the superior ability of the LogNormMix-Net multi-class recipient classification approach over the alternative multi-label classification approach. Then we address the event generation component of our second experimental aim in section \ref{TPP-gen}, by evaluating the realism of the LogNormMix-Net generated content.
\subsubsection{Model goodness of fit and predictive ability}\label{goodness-of-fit}
Since the temporal and sender modules are identical across both the LogNormMix-Net and LogNormMix-BC models, it is reasonable to expect them to be able to achieve similar performance for inter-arrival time and sender prediction. We include results for these modules to validate that adding the recipient prediction task doesn't unreasonably degrade performance on those two original tasks. The main goal of this section, though, is to demonstrate that the LogNormMix-Net can learn the relationship between senders and the recipient sets they communicate with.

The results for time prediction and sender/recipient prediction on the test set are summarized in Tables \ref{tab:timepredict} and \ref{tab:participantpredict} respectively. We see that the multi-label classification approach, LNM-BC, performs similarly to the original LogNormMix model on time and sender prediction, but is completely unable to predict the next recipient set correctly given the sender. On the other hand, the LogNormMix-Net is able to correctly predict the next recipient set given the sender with an accuracy of 33.6\% and 40.8\% on the Enron and EU datasets respectively, and with top-3 accuracy of 61\% and 71\%. Results for the event prediction NLL show a similar pattern and are included in Table \ref{tab:NLL} in Appendix \ref{appendix-results}.  

Tables \ref{tab:timepredict} and \ref{tab:participantpredict} establish that while the LogNormMix-Net and LogNormMix-BC models have similar ability to model the temporal and sender distribution, only the LogNormMix-Net multi-class classification approach to recipient modelling can capture the relationship between senders and the recipient sets they communicate with. Further evidence for this can be found in Table~\ref{tab:enron-small-54} in the next section. Together, these results validate our first experimental aim.

\begin{table}[t]
\caption{Test set event Time predictions.}\label{tab:timepredict}
    \begin{tabular}{l  | c  c | c  c }
    \multirow{2}{*}{} & \multicolumn{2}{c}{\textbf{Enron}} & \multicolumn{2}{c}{\textbf{EU}} \\
      \textbf{Model} & RMSE (hrs) & MAE  & RMSE & MAE \\
      \hline 
      THP & 11.5 & -  & & - \\
      LogNormMix & \textbf{5.9} & \textbf{3.0} & 3.63& 2.41\\
      \hline
      LNM-Net & 6.1 & 5.3 & \textbf{2.96} & \textbf{0.93}\\
      LNM+BC  & \textbf{5.9} & 3.6  & 7.72  &  6.38 \\
    \end{tabular}
\end{table}

\begin{table}[t]
     \caption{Test set participant prediction accuracy.}\label{tab:participantpredict}
    \begin{tabular}{l|ccc|ccc}
    \multirow{2}{*}{} & \multicolumn{3}{c}{\textbf{Enron}} & \multicolumn{3}{c}{\textbf{EU}}\\
     \textbf{} & Sender & Recip & Recip & Sender & Recip & Recip \\
      \textbf{Model} & Top-1 & Top-1 & Top-3 & Top-1 & Top-1 & Top-3 \\
      \hline 
      THP &  9.7\% & - & - & & - & -\\
      LNM & 17.5\% & - & - &13.6\% & - & -\\
      \hline
      LNM-Net & \textbf{19.4\%} & \textbf{33.6\%} & \textbf{61.1\%} & \textbf{22.4\%} & \textbf{40.8\%} & \textbf{71.3\%}\\
      LNM+BC  & 17.7\% & 0.05\% & & 17.16\% & 0.00\% &\\
    \end{tabular}
\end{table}

\subsubsection{Evaluating generated content}\label{TPP-gen}
In table~\ref{tab:enron-small-54} we use the Earth Mover's Distance (EMD) to evaluate the fidelity of the generated data to the ground truth training data. Results show that the scores for the 3 models are reasonably balanced across the two datasets for all measures except recipient set indegree. For the Enron dataset, the parametric Hawkes TPP model of Fox et al. \cite{fox2016enron} was the worst of the 3 models at reproducing the time-delta distribution, but the best at reproducing the sender distribution and the temporal seasonality measures (ie. proportion of events per hour and per day). For the EU email dataset, LogNormMix-Net was the best for all measures except the day of week seasonality. For the recipient set indegree measure, neither the Fox Hawkes TPP nor the LogNormMix-BC were able to reproduce the ground truth distribution. Analysis showed that around 33\% of the time, the Hawkes process TPP model sampled recipient sets that had not been seen in the training set, which we therefore consider invalid, and for the LogNormMix-BC 99\% of the recipient sets were invalid. The LogNormMix-Net, on the other hand, was able to reproduce the proportion of emails received by each recipient set in its generated event streams, whilst maintaining strong performance across all the other measures, including time-delta and sender proportions. 

\begin{table*}[t]
    \caption{EMD between real and simulated data on the Enron and EU email corpora. The values in the columns are the means of the specified variables over the 100 trials, with corresponding standard deviations given in parenthesis.}\label{tab:enron-small-54}
    \begin{tabular}{l  | c | c | c | c | c | c}
      \hline 
      \textbf{Model} & Dataset & Time  & Hour & Day & Sender  & Recip set \\
     & & deltas & of day & of week & outdegree & indegree \\
      \hline 
      Fox Hawkes TPP & Enron & 1.86 (0.09) & \bf{1.04 (0.09)}& \bf{0.87 (0.04)} & \bf{1.01 (0.24)} & 33.80 (1.22) \\
      LogNormMix-BC  & Enron & \bf{0.42 (0.05)} & 1.81 (0.12) & 1.26(0.08) &  5.33 (0.22) &  669.14 (49.15) \\
      LogNormMix-Net  & Enron &  0.61 (0.10) & 2.18 (0.11)& 0.92 (0.06)& 4.61 (0.40)  & \bf{6.49 (0.47)} \\
      \hline 
      Fox Hawkes TPP & EU & 0.39 (0.10) & 3.08 (0.10) & 0.57 (0.26) & 8.95 (0.88) & 34.02 (0.85) \\
      LogNormMix-BC  & EU & 0.27 (0.22) & 2.57 (0.23) & \bf{0.40 (0.12)}& 3.06 (0.20) & 2186.59 (943.94)  \\
      LogNormMix-Net  & EU & \bf{0.24 (0.09) }& \bf{2.35 (0.17) }& 0.49 (0.12) & \bf{0.64  (0.14)} &  \bf{3.81  (0.82)} \\
      \hline 

    \end{tabular}
\end{table*}

\begin{figure}[htb]
    \centering
    \includegraphics[width=\columnwidth]{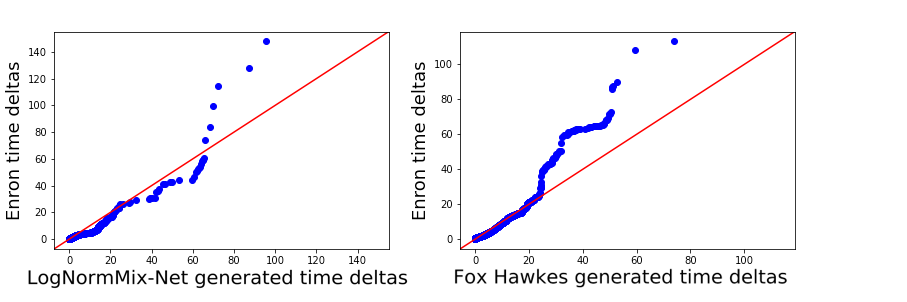}
    \caption{Q-Q plot comparison of inter-arrival times generated by LogNormMix-Net (left) Vs the Fox Hawkes TPP (right) on Enron data.}
    \label{hist2}
\end{figure}

Providing a visual representation of the results in Table~\ref{tab:enron-small-54} provides better intuition for whether these results produce realistic enough traffic for the purpose of cyber deception. The Q-Q plots in Figure \ref{hist2} indicate that both models do a reasonable job of reproducing the ground truth distribution for the time deltas. The LogNormMix-Net has a slight tendency to under-generate (generate too few points). On the other hand, the Fox Hawkes parametric TPP tended to substantially over-generate (generate too many points). In order to mitigate this, we placed thresholds on the Fox Hawkes generative model to curb the number of events generated. Despite this, we see in Figure \ref{hist2} that the generated time-deltas in general were still smaller than the ground truth. 

\begin{figure}[htb]
    \centering
    \includegraphics[width=\columnwidth]{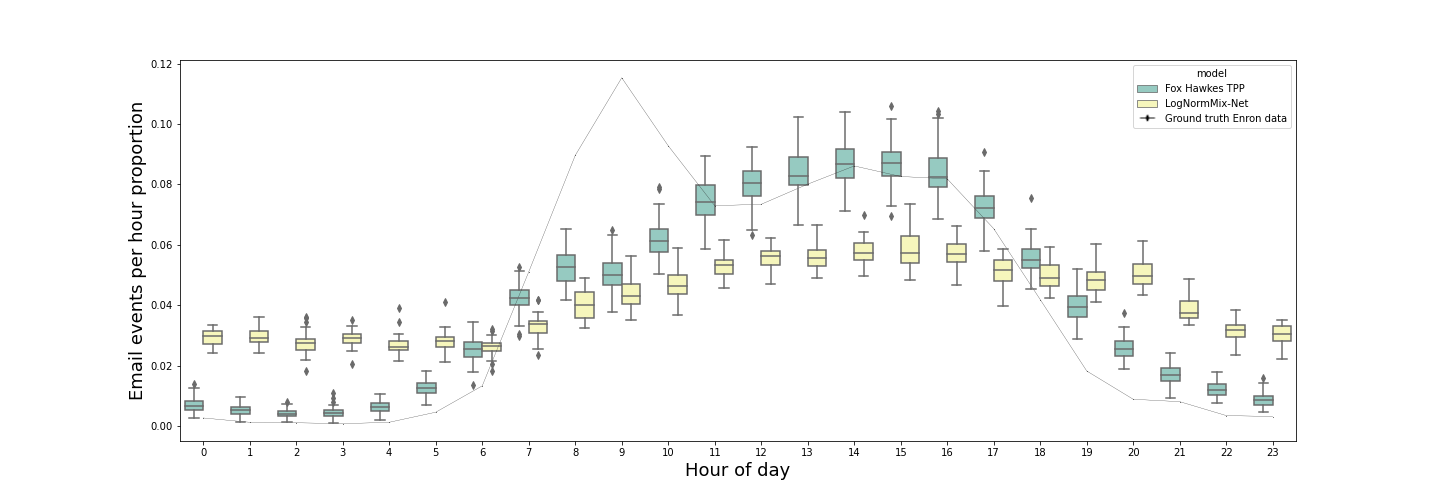}
    \caption{Proportion of emails generated by hour of the day by LogNormMix-Net Vs the Fox Hawkes TPP model.}
    \label{hour-hist}
\end{figure}

\begin{figure}[htb]
    \centering
    \includegraphics[width=\columnwidth]{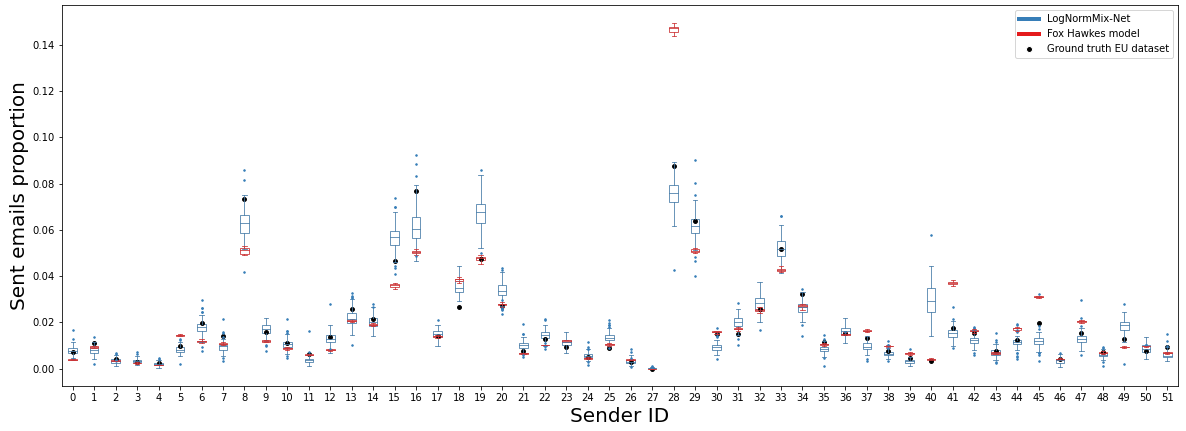}
    \caption{Proportion of emails sent per user in LogNormMix-Net generated data Vs Fox Hawkes TPP generated data on the EU email dataset.}
    \label{hist1}
\end{figure}

\begin{figure}[htb]
    \includegraphics[width=\columnwidth]{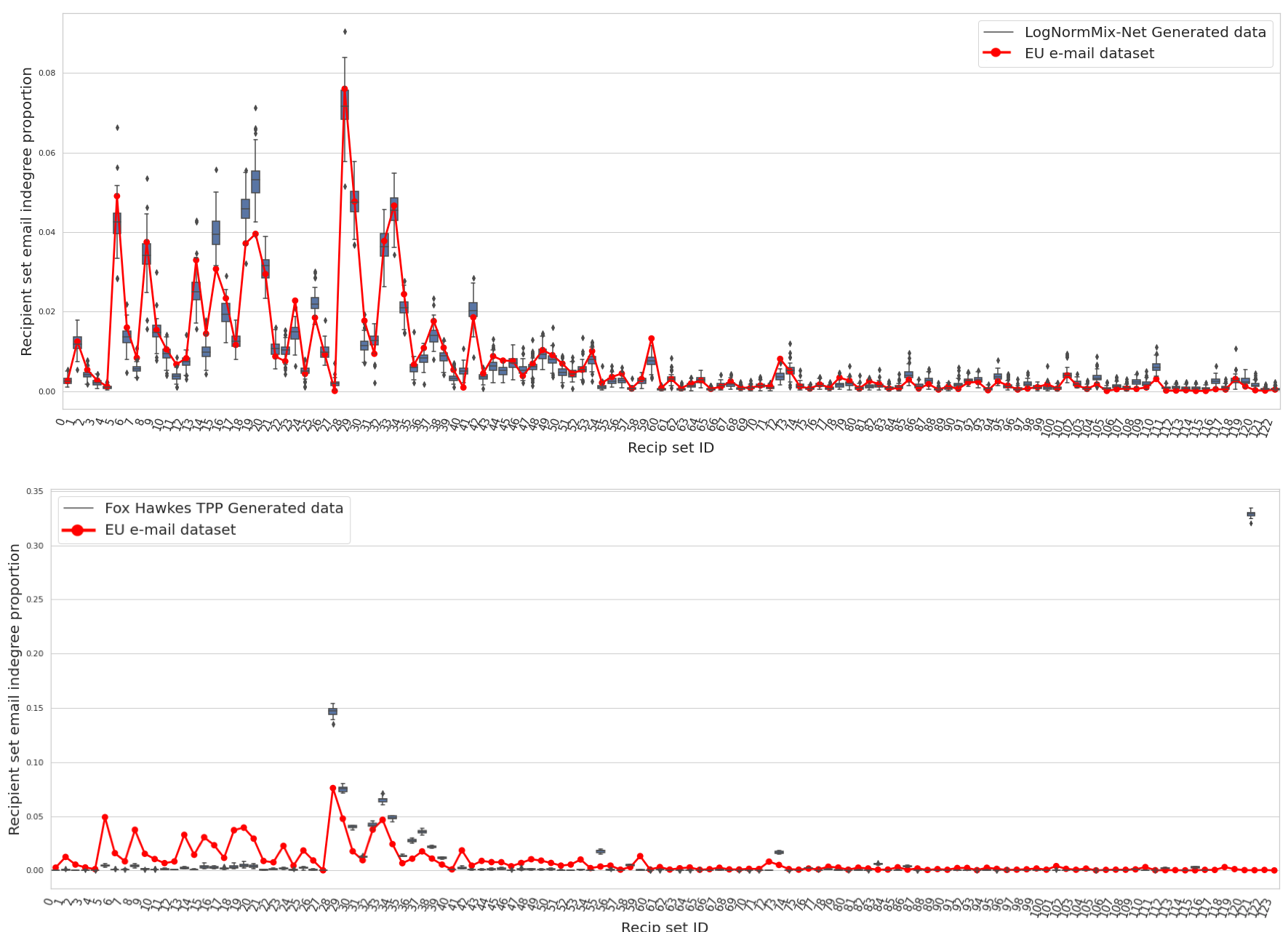}
    \caption{Received emails proportion by ID for generated LogNormMix-Net (above) Vs Fox Hawkes TPP (below).}
    \label{reciphistogram}
\end{figure}

Figure \ref{hour-hist} shows that both the LogNormMix-Net and Fox Hawkes TPP model are able to capture some of the seasonality effect, though it is smoothed/dampened for the LogNormMix-Net. Figure \ref{hist1} shows that the Fox Hawkes TPP model is able to reproduce the sender distribution reasonably well, but we see in Figure \ref{reciphistogram} that the Fox Hawkes model is unable to reproduce the ground truth recipient distribution. For the purpose of visualisation, all recipient sets generated by the Fox Hawkes model that were not seen in the training data were assigned to Recipient Set ID 123. We see in in Figure \ref{reciphistogram} that ID 123 accounts for roughly one third of generated recipient sets by the Fox Hawkes model, and as a result, the majority of the legitimate recipient set IDs were under-sampled by the model. One the other hand, Figures \ref{hist1} and \ref{reciphistogram} show that the LogNormMix-Net is able to reproduce both the sender and recipient set distributions from the EU corpus training data respectively, with the true sender proportion overlapping the boxplot IQR of generated content for the majority of sender/recipient IDs. These results demonstrate that our model is able to model the temporal and sender characteristics as well as the other TPP models, but in addition, supports the essential feature of modelling the recipients, which the other models cannot do.


\subsection{Multi-party email thread generation}\label{thread-gen}
Sections \ref{sub_pers} and \ref{text-gen-results} below are devoted to addressing the remaining component of our second experimental aim. Namely, to evaluate the realism of the generated conversation threads.
\subsubsection{Subject personalisation}\label{sub_pers}
The goal of extracting keywords for each user is to personalise the topics of generated threads. Examples from the generated subjects of user IDs 4 and 15 are shown in Table~\ref{genkeysub}. User ID 4's emails have a theme of communications and reporting, with the word "update" being used in 22\% of the generated subjects, including various versions of recurring "Weekly Update" themed subjects, and other recurring updates on high profile portfolios like the "Pipeline" project, while ID 15's emails have a theme of scheduling meetings, interviews, and other recruitment tasks. An additional example of a user who may be in sales or legal is shown in Figure \ref{sales-guy} in Appendix \ref{appendix-text1}.

\begin{table*}[htb]
\caption{Keywords and resulting generated subjects for user IDs 4 and 15.}
\label{genkeysub}
\begin{tabular}{lll}
\multicolumn{1}{c|}{\textbf{ID 4}}                                                                                                                             & \multicolumn{2}{c}{\textbf{ID 15}}                                                                                                                                                                             \\ \hline
    \multicolumn{1}{l|}{\shortstack{Keywords: Update, Meeting, Bullets, Weekly,\\\quad Capacity, Pipeline, Storage, Project, Revised, List}} 
    & \multicolumn{1}{l|}{\shortstack{Keywords: Meeting, Conference, Risk, Resume,\\ Visit, Energy, Research, Power, Model, Summer}} & Subjects starting "Conference Call"            \\ \hline
\multicolumn{3}{c}{\textbf{Generated Subjects}}                                                                                                                                                                                                                                                                                                                                  \\ \hline
\multicolumn{1}{l|}{Top 10 Things You Wish Your ISP Would Notice.}                                                                                     & \multicolumn{1}{l|}{Risk Management Simulation-Please review..}                                                                                           & \multicolumn{1}{l}{.. Monday, April 19, 2001.}                          \\
\multicolumn{1}{l|}{Pipeline News: August 26, 2001.}                                                                                                           & \multicolumn{1}{l|}{Summer and Fall Schedule, November.}                                                                                                  & \multicolumn{1}{l}{\shortstack{.. Scheduled for 6pm on Tuesday,\\June 9th.}}           \\
\multicolumn{1}{l|}{Pipeline Summary for October 11, 2001.}                                                                                                    & \multicolumn{1}{l|}{Summer Intern Information.}                                                                                                           & \multicolumn{1}{l}{.. on Tuesday, November 11th.}                       \\
\multicolumn{1}{l|}{Meeting to discuss Team Selection -Reply.}                                                                                                 & \multicolumn{1}{l|}{Summer Associate Candidate - Angela Davis.}                                                                                           & \multicolumn{1}{l}{.. re: Teams.}                                       \\
\multicolumn{1}{l|}{Update on California Electricity Market.}                                                                                                  & \multicolumn{1}{l|}{Resume for Jeff Skilling.}                                                                                                            & \multicolumn{1}{l}{.. .}                                                \\
\multicolumn{1}{l|}{Bullets 09/02/01.}                                                                                                                         & \multicolumn{1}{l|}{Resume : Your Input Required.}                                                                                                        & \multicolumn{1}{l}{.. : Trading Floor}                                   \\
\multicolumn{1}{l|}{Capacity Matrix Update.}                                                                                                                   & \multicolumn{1}{l|}{Power Point Presentation on Credit Risk.}                                                                                             & \multicolumn{1}{l}{.. Cancelled.}                                       \\
\multicolumn{1}{l|}{Capacity Report.}                                                                                                                          & \multicolumn{1}{l|}{Energy Analysis - New Issue.}                                                                                                         & \multicolumn{1}{l}{.. to discuss the P\&D Program.}                     \\
\multicolumn{1}{l|}{Weekly Update from the Office of the Chairman.}                                                                                            & \multicolumn{1}{l|}{Risk Systems Update for December 11th.}                                                                                               & \multicolumn{1}{l}{\shortstack{.. to Discuss California/West \\\quad Wholesale Activities.}} \\

\multicolumn{1}{l|}{Weekly Updates: Energy, Environment, and Weather.}                                                                                         & \multicolumn{1}{l|}{Visit to  Portland  - July 18.}                                                                                                       &                                                     \\
\multicolumn{1}{l|}{Weekly Update on Power Markets \& Energy Market.}                                                                                          & \multicolumn{1}{l|}{Visit to Weather Desk.}                                                                                                               &                                                     \\
\multicolumn{1}{l|}{Weekly Update - RTO Week - Summary of Comments.}                                                                                          & \multicolumn{1}{l|}{Model Review Meeting - June 9, 2001.}                                                                                                 &                                                    
\end{tabular}
\end{table*}

These results support our claim that this keyword generation approach can be applied to differentiate sender’s email topics from one another, and give an indication of their role in a network. 

Repetition was not a problem in the generated subjects, even among subjects where the first couple of words were identical. For example, user ID 15 had many subjects being generated about meetings and conference calls, and it was observed that the subject generation model generated different dates and/or descriptors for each subject, as shown in the right-most column of Table \ref{genkeysub}.

\subsubsection{Email thread generation}\label{text-gen-results}
\begin{figure}
\includegraphics[width=\columnwidth]{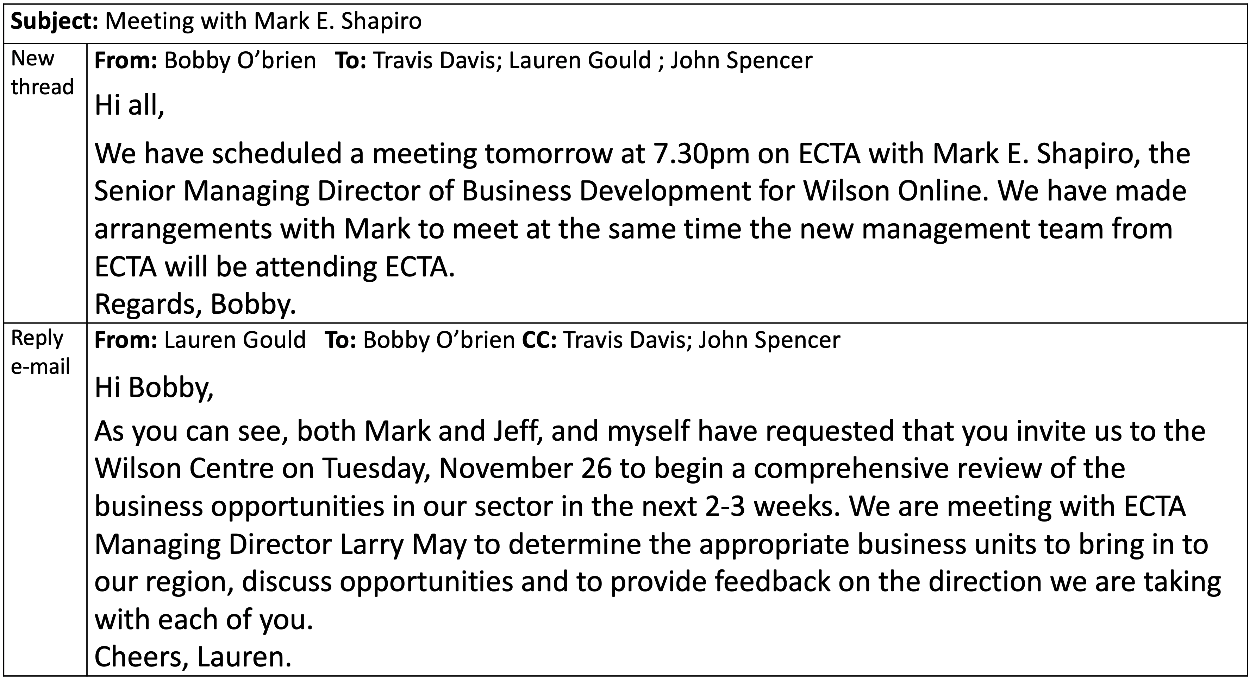}
\caption{Example of a generated email/reply pair.}\label{GPT2-thread}
\end{figure}

\begin{figure}
\includegraphics[width=\columnwidth]{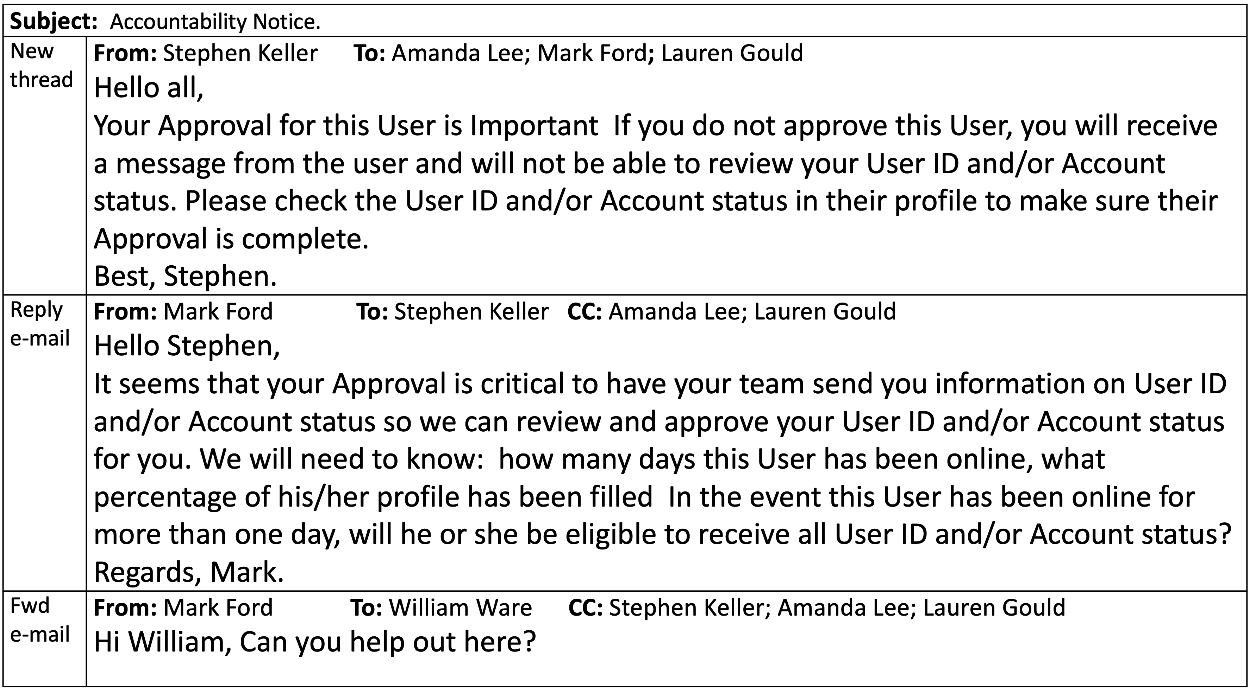}
\caption{An example generated  thread containing the original email, a reply and a fwd email.}\label{GPT2-thread3}
\end{figure}

\begin{figure}
\includegraphics[width=\columnwidth]{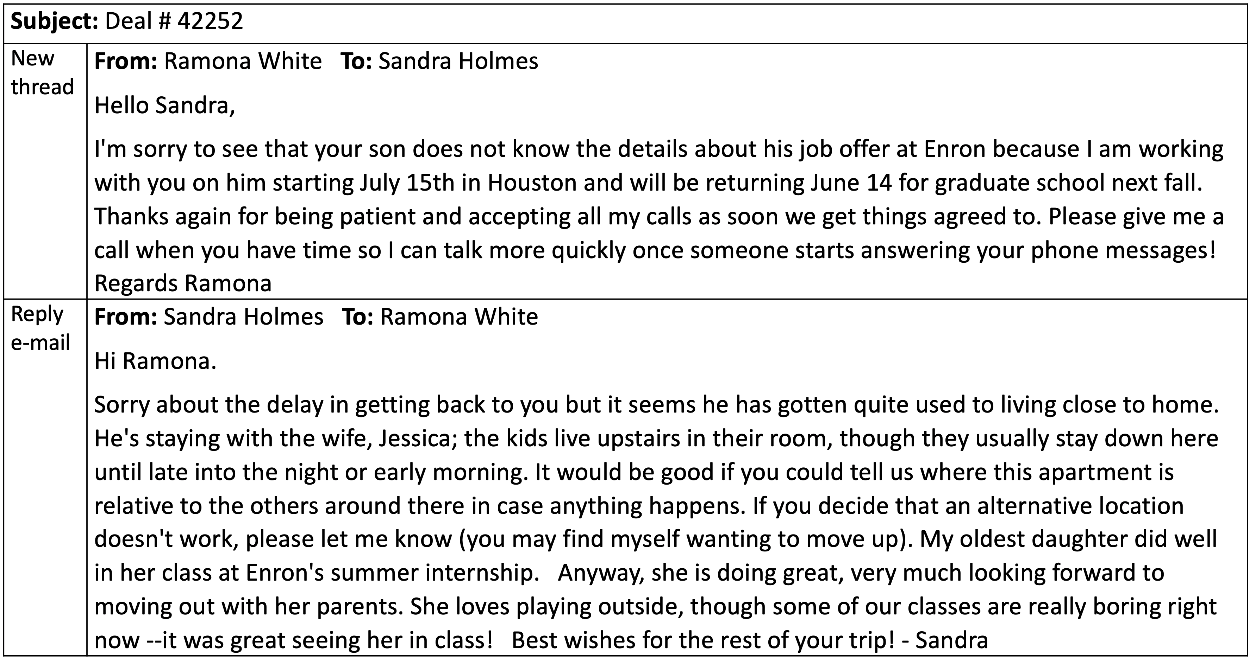}
\caption{Example generated email/reply pair from the model trained using conversation thread details.}\label{GPT2-threadtrain}
\end{figure}

Example email threads generated using our GPT-2-based email body generation models are shown in Figures \ref{GPT2-thread}-\ref{GPT2-threadtrain}. The threads in Figures \ref{GPT2-thread} and \ref{GPT2-thread3} were generated using a GPT-2 email body generation model whose training dataset treated each email body independently of any previous emails in the thread. The example thread in Figure \ref{GPT2-threadtrain} was generated by a different GPT-2 model where the training data included email thread information. Overall, the authors felt the replies generated using the second model (Figure~\ref{GPT2-threadtrain}) were typically more directly relevant and arguably created a more coherent thread than the first model.

The coherence and cohesion scores for each of the email-reply pairs in our generated examples, as well as the average of the pairs in the Enron test comparison dataset are show in Table \ref{tab:coherence}. These show that our model is able to produce similar scores to the Enron dataset that it was fine-tuned on. Figure \ref{GPT2-thread3} shows an interesting phenomenon we observed in both our generated email threads as well as in the Enron dataset. The first email and it's reply have a high similarity score of 0.79, but the second and third emails have a very low similarity score of 0.006. The low score is a result of the third email, a forward of the thread to a new recipient to elicit help. This occurrence of an email pair with low similarity scores in a thread with otherwise high similarity was observed in 4 of the 10 Enron email threads in our test dataset. It was seen to happen when a reply or forward email comprised only a short question/statement without context, e.g. "Any progress?" or "I'll take care of that". We include an example of this from Enron in Appendix \ref{appendix-enron-thread}, Figure \ref{tab:enron}. From the point of view of the authors, all the above-mentioned threads in the Enron dataset exhibiting this phenomenon were perfectly coherent and cohesive communications, thus highlighting an opportunity for the development of improved metrics for evaluating communication thread realism.

The results presented in sections \ref{TPP-gen} and \ref{thread-gen} show that our end-to-end email simulation framework can generate convincing content, appropriate for application in cyber deception. Unlike the Fox Hawkes TPP and the LogNormMix-BC model, our LogNormMix-Net model is able to generate event sequences that reproduce the underlying temporal, sender and recipient distributions of the ground truth training data (Table~\ref{tab:enron-small-54} and Figures \ref{hist2} - \ref{reciphistogram}). Secondly, our email thread generation approach can generate communication threads with similar coherence and cohesion scores to the real-life Enron dataset that it was fine-tuned on (Table \ref{tab:coherence}), and with subjects personalised to the job roles and/or interests of the person initiating the communication thread (Table \ref{genkeysub}). These combined results therefore validate our second experimental aim.

\begin{table}[htb]
\caption{Coherence and cohesion scores for our generated examples Vs Enron emails average scores.}
\label{tab:coherence}
\begin{tabular}{l|c|c|c|c|c|c}
\textbf{} &
  \textbf{Enron} & \scellb{Gen.}{Corpus} & \scellb{Thread 1}{(Fig. 8)} & \multicolumn{2}{c|}{\scellb{Thread 2}{(Fig. 9)}} & \scellb{Thread 3}{(Fig. 10)} \\ \hline
  \textit{Emails} & \multicolumn{2}{c|}{\textit{Avg. of}} & \textit{ } & \textit{ } & \textit{ } & \textit{ } \\ 
  \textit{Compared} & \multicolumn{2}{c|}{\textit{All Pairs}} & \textit{ 1\&2} & \textit{1\&2 } & \textit{2\&3 } & \textit{1\&2 } \\ \hline
\textbf{Coherence} &
  \multicolumn{1}{c|}{0.31} &
  0.43 &
  0.51 &
  \multicolumn{1}{c|}{0.79} &
  0.006 &
  0.43 \\ \hline
\textbf{Cohesion} &
  \multicolumn{1}{c|}{0.13} &
  0.17 &
  0.20 &
  \multicolumn{1}{c|}{0.29} &
  0.04 &
  0.14
\end{tabular}

\end{table}

\subsection{Scalability}
In order to be a useful tool for cyber deception, our framework must be scalable, and have the capacity to automatically generate as many events and as much message content as required without human intervention. As a minimum requirement, our deployed model must be able to generate events in real time. Running on a single Nvidia Tesla P100 SXM2 (16GB) GPU and an Intel Xeon 14-core E5-2690 v4, it takes 13 minutes to generate a week’s worth of events ($\approx300$ emails) for the EU Email dataset network of 52 people. The text generation component accounts for more than 99\% of the time, with the TPP event generation taking less than 7 seconds for the 300 events. Since our code is not written to use multi-GPU processing at the current time, it could be further improved by generating the text for separate email threads on different GPUs.

\section{Discussion}
This section is organised into the following four topics: Models, Applications, Generated Content, and Communication. In the following we discuss the Models and Generated Content, and the discussion of Applications and Communication are presented in Appendix \ref{discussion-applications} and \ref{discussion-communication}. 

\subsection{Models}
\textbf{Multi-class Vs multi-label recipient classification}
In this work we have demonstrated the superior ability of the multi-class classification approach to model recipient set selection in DM communications. A further and significant advantage is that the LogNormMix-Net presents a single neural TPP architecture to simulate both unicast data, seen for example in WiFi network traffic, and multi-cast data, seen in DMs modelled in this work. Multi-class recipient selection is the natural choice for modeling unicast communications, whereas multi-label classification recipient selection does not make sense, since only one recipient is involved for each event. 

A drawback of multi-class classification approaches is lack of scalability, thus requiring infrequent recipient sets to be discarded from training data. This is not necessarily a problem for cyber deception, but a multi-label classification approach may be preferred for very large DMNs, or when there is a preference to use a training dataset without needing to refine it. A potential way to get around the limitations of the multi-label binary classification approach is the use of classifier chains, an extension which instead builds a chain of classifiers, where label information is passed between the classifiers. 

\textbf{Comparison with the Fox Hawkes TPP}
The elegant multivariate parametric Hawkes TPP~\cite{fox2016enron} models the senders and event time of email communications. Unlike the LogNormMix-Net, it does not model recipients, and is therefore not a network TPP in the sense that we consider in this paper. We found that trivially extending the author's event generation algorithm to multi-cast events caused the model to generate far too many events. This could be because adding additional recipients caused too many excitation events to be triggered by the Hawkes Process. In addition to this, the sampling approach was also unable to reproduce the ground truth recipient set distribution (cf Figure \ref{reciphistogram}). Our LogNormMix-Net model on the other hand is able to model event times and senders as well as the Fox Hawkes TPP, but additionally supports some essential features of DM modelling such as recipient modelling, which are not supported by current TPP models. For these reasons, the Fox Hawkes TPP's generated content was found to be less realistic than the LogNorMix-Net. Another drawback of this approach is that it takes roughly 25 times as long to train the Fox Hawkes TPP than the LogNormMix-Net. Lastly, the sampling algorithm from the Fox paper \cite{fox2016enron} proposes to generate events from their Hawkes Process in layers, which means events would need to be pre-generated before deployment time, rather than generated in real-time. This makes it unsuitable for deployment in a cyber deception system that requires real time event generation.

\textbf{Task prioritization in multi-task learning}
Our LogNormMix-Net can be thought of as a heterogeneous multi-task learning framework. We found that the temporal module trains a lot faster than the classification modules, and training until the minimisation of the the combined NLL loss resulted in over-fitting of the temporal module. We note that the balancing of multiple tasks to optimise performance is an ongoing research problem in multi-task learning. In our case an alternate training strategy was implemented. The most basic approach of weighting the loss terms was not sufficient. Instead we trained the tasks in stages, and tracked the mean absolute error (MAE) and the root mean squared error (RMSE) of the inter-event time prediction during training to help inform this process.
\begin{enumerate}
    \item Stage 1: we let all tasks train for a number of epochs until the RMSE of time prediction on the validation set begins to degrade.
    \item Stage 2: we freeze the parameters associated with the temporal module (ie. the Lognormal mixture linear layer and the metadata embedding) and the RNN that embeds the history. We then continue to train the sender and recipient prediction modules.
    \item Stage 3: (optional) additionally freeze the sender module and continue to train the recipient module.
\end{enumerate}

\textbf{Independence assumption between marks and inter-arrival times}
Our TPP approach models the event time and participants as conditionally independent given the history, since this independence assumption leads to significant benefits for the speed of event generation at deployment time. If instead the time was modeled to be conditionally dependent on the participants, then the observed sender, receivers, and timestamp of each event would be generated by selecting the sender–receiver-set pair with the smallest time delta~\cite{Snijders96}. This approach requires a candidate recipient set and time delta to be generated for every sender in the network for every event, though, which increases the event generation time.

Our LogNormMix-Net can be modified to jointly model time and marks by mimicking the Neural Hawkes \cite{mei2017neural} architecture and directly modeling the distribution of $n$ inter-event times separately, using a shared RNN to process the history. However, this requires each sender to have a sufficient number of events to learn an individual temporal mixture model. 

\textbf{Graph Neural Network (GNN) based approaches for TPPs}
Multi-dimensional TPPs are inherently graph based, and are a good way to combine the temporal and graph aspects of network communication modelling. Recent work of Zhang and Yan \cite{ijcai2021-469} models TPPs using a GNN-based architecture. Their work doesn't model network communication events (ie. they don’t model recipients), so would need to be extended to fit our use case. Their model also doesn't have closed-form likelihood or sampling, which were the two key reasons we chose the intensity-free LogNormMix approach. 

Temporal Knowledge Graphs (TKGs) are another interesting graph-based approach. Han et al. \cite{han2020graph} designed a Hawkes Process inspired TKG approach for event forecasting. An event generation algorithm can be written for their model, however it would be a lot slower to generate events at deployment time for the same reasons as described in the paragraph above that discusses our choice to model marks and inter-arrival times independently. Additionally, there is no closed form expression for event generation for their model.

Developing GNN- or TKG-based TPP approaches for multi-cast network communication modelling, with efficient likelihood computation and event sampling is an interesting problem for future work.

\subsection{Generated Content}
\textbf{Privacy of generated content}
Approaches for fine-tuning large language generation models like GPT-2 under Differential privacy (DP) have recently been proposed which show promising results towards preventing unintended memorization of downstream data\cite{li2021large, yu2021differentially}. In cases where an organisation wants to fine-tune GPT-2 on their own email corpus for use in our framework, fine-tuning under DP as in \cite{li2021large} could be utilized. 

For the second component of our framework, our TPP-based event generation model is capable of learning some behavioural information and interactions patterns from the training data. DP for TPP event generation is therefore an interesting open problem for future work.

Aside from DP, there is a spectrum of anonymisation that can be applied. By using fake names in the deployed generative model instead of the real names of people from the corporate network, the risk of interaction patterns learned by the TPP model being associated with real people would be greatly reduced. The likeliest source of real confidential information is the content of real messages. So open source message corpora could be used to fine-tune the language generation model, while the organisation's real data can still be used to train the TPP event model. At the most cautious end of the spectrum, independent open source data could be used to train both the event generation and text content generation models, at the cost of some realism.

\textbf{Evaluating the realism of generated communication threads}
We’ve measured coherence and cohesion in this paper because they are popular in the literature, and we posit that they are a necessary requirement for realism, but we acknowledge that it's not a universal answer the question of “is this content realistic?”. The overall question of long-range realism over a communication thread is quite hard. Human perception of language is very sophisticated and quite subtle, and for this reason user studies may be employed to evaluate the realism of generated text ~\cite{clark2021all,shabtai2016honeytoken}. Our team are working towards a user study which compares human evaluation versus the coherence and cohesion metrics used in this work. One limitation of applying cohesion and coherence as in Karuna et al. \cite{karuna-etal-2018-enhancing} to communication threads that we identified in this work is that they both penalize responses that don't repeat any context of previous message (cf email 3 in Figure \ref{GPT2-thread3}). Since such responses are common, for example in forwarded emails that loop in an additional person, a task for future work is to come up with more sophisticated metrics that compensate for such communication patterns.

\section{Conclusion} In this work we present a framework to automate the generation of email and instant messaging-style group communications at scale, for use in cyber deception. We introduce the LogNormMix-Net TPP which is a single neural TPP architecture capable of simulating both unicast and multi-cast data that is flexible, and benefits from closed form likelihood and sampling. We demonstrate the superior ability of the LogNormMix-Net to model recipient set selection over existing recipient modeling approaches, and validate that generated content captures the underlying dynamics of real world email network traffic datasets, essential for applications in deception. We demonstrate the ability of fine-tuned, pre-trained language models to generate coherent, topical, multi-party conversation threads, and then propose a framework for uniting our generated TPP and language content for real-time simulation of an email server. Such content could be used as a honeypot or to add realism to larger deceptive environments.

\bibliographystyle{IEEEtran}

\appendices
\section{Further details on the LogNormMix-Net and traffic generation}
\label{appendix-results}
\subsection{LogNormMix-Net architecture}\label{lognormmix-arch}
\begin{figure}[htb]
\includegraphics[width=\columnwidth]{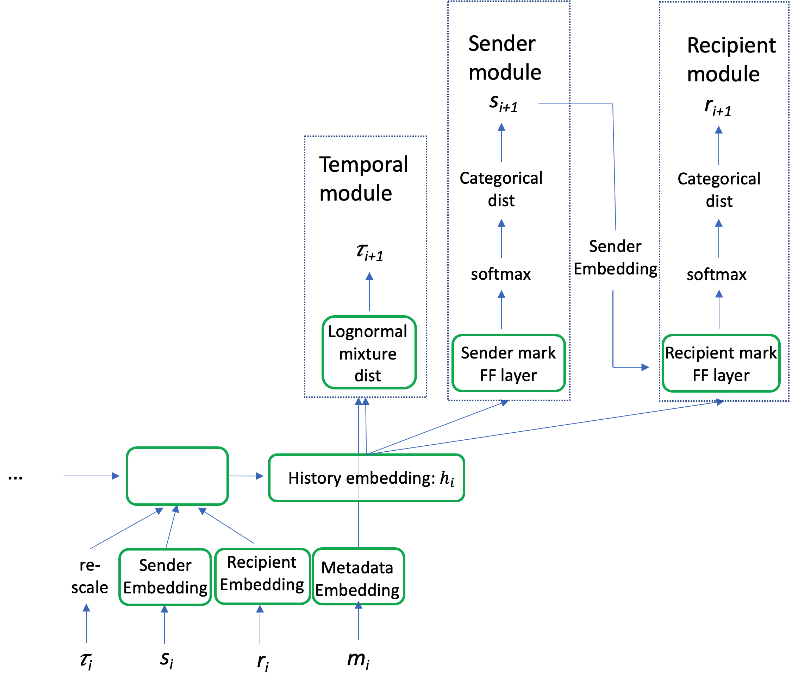}
\caption{LogNormMix-Net conditional architecture. }\label{conditionalarch}
\end{figure}

Figure \ref{conditionalarch} shows the LogNormMix-Net architecture overview.

\subsection{NLL results for event prediction experiments}
Table \ref{tab:NLL} shows the NLL results for our event prediction experiments from Section \ref{results-section}. We see that the LogNormMix-Net (denoted LMN-Net) gets the lowest NLL for the sender and recipient prediction task, with the original LogNormMix getting the lowest for time prediction (though it has no recipient module and thus is only solving 2 tasks instead of 3).

\begin{table}[htb]
    \caption{NLL by task and dataset}\label{tab:NLL}
  \begin{center}
    \begin{tabular}{l  | c c c  | c c c }
      \multirow{2}{*}{\textbf{\newline \newline Model}} & \multicolumn{3}{c}{\textbf{Enron}} & \multicolumn{3}{c}{\textbf{EU}}\\
        & \textbf{Time} & \textbf{Sender} & \textbf{Recip}  & \textbf{Time} & \textbf{Sender} & \textbf{Recip}  \\
        & \textbf{NLL} & \textbf{NLL} & \textbf{NLL} & \textbf{NLL} & \textbf{NLL} & \textbf{NLL} \\
      \hline 
      LogNormMix & \textbf{74.4} & 361.0  & - & \textbf{-68.2} & 980.3   & -  \\
      LMN-Net & 74.7 & \textbf{340.5} & \textbf{447.8}  &-63.1 &  \textbf{857.4} &  \textbf{1004.2} \\
      LMN-BC  & 78.1 & 347.6 & 625.0   &-2.1 &  1002.8 &  6851.2 \\
    \end{tabular}
  \end{center}
\end{table}

\section{Further details on email content generation}
\subsection{Selecting email communication type}\label{FirstAppendix}
We consider 3 types of communications: \\
    i) Starting a new conversation thread\\
    ii) "reply": Replying in an existing conversation thread\\
    iii) "fwd": Forwarding the existing communication to a new recipient (note: although unrestricted forwarding of communications isn't possible in platforms like Slack, it does have the functionality to share a message from a public channel or a channel the recipient is a member of, to a private message to said recipient). \\

For each sender, we aim to ensure the proportions of each communication type (new-thread, reply, fwd) in the rolling simulated content resembles the proportions of that sender from the training data. The pseudocode for email type selection is provided below.

\textit{Pseudocode for selecting email type given the sender and recipients.}\label{email-type} \\
\textbf{if} there's an existing active email thread between those participants, and the proportion of "reply" type emails sent by the sender in the past "recent" time period (eg. 2 months) of simulated events is less than 1.1 times the training dataset proportion:\\
    \-\hspace{0.25cm} email type = "reply"\\
\textbf{else if } {there's an existing active thread between a subset of those participants, and the proportion of fwd messages sent by that sender in the past "recent" time period (eg. 2 months) of simulated events is less than 1.1 times the training dataset proportion:}\\
    \-\hspace{0.25cm} email type = "fwd"\\
\textbf{else: }\\
    \-\hspace{0.25cm} email type = "new thread". \\

\subsection{email thread generation}\label{email thread algo}
Here we describe the steps to generate an email thread given the sender and recipients.\\
\textit{email thread generation}\label{emailgen} 
\begin{enumerate}
    \item[a:] \underline{Sample the communication type} (new thread, reply, fwd) based on that sender's proportions from the training data, according to Algorithm \ref{email-type} in Appendix \ref{FirstAppendix}.
    \item[b:] \underline{Generate the email Subject} to begin a new thread \underline{or} \underline{sample an existing thread} to respond to (depending on the communication type sampled in step a): \\
    \\
        \textbf{if} "reply"/"fwd" communication: \\
        \-\hspace{0.25cm} i) sample a conversation thread to respond to, that is: \\
        \-\hspace{0.5cm} that is: 
        \-\hspace{0.5cm} - recently active (eg. in the last week) \\
        \-\hspace{0.5cm} - appropriate for the recipients selected in step\\
        \-\hspace{0.8cm} \textit{a}, and communication type.\\
        \-\hspace{0.25cm} ii) inherit the thread ID from the selected thread.\\
        \textbf{else if} "new thread": \\
        \-\hspace{0.25cm} i) generate email Subject: \\
        \-\hspace{0.5cm} - sample from the sender's topic keywords\\
        \-\hspace{0.5cm} - feed that topic into the subject generation model \\
        \-\hspace{0.8cm} as the prompt for the generation. \\
        \-\hspace{0.25cm} ii) generate a new ID for this new email thread.
    \item[c:] \underline{Generate email body text}:\\
    i) Generate a greeting: randomly sample a greeting \\
    \-\hspace{0.25cm} from a canned list, and append the recipient name(s) \\
    \-\hspace{0.25cm} to complete the greeting.\\ 
    ii) Generate the body message text: \\
    \-\hspace{0.25cm} \textbf{if} "fwd" communication:\\
            \-\hspace{0.5cm} - randomly sample the body text message from a\\
            \-\hspace{0.7cm}  canned list of responses for "forward"-type emails, \\
    \textbf{else:}\\
            \-\hspace{0.5cm} use the email subject and existing email thread \\
            \-\hspace{0.5cm} (if any) to seed the email body generation model.\\
    iii) Generate the salutation: randomly sample a salutation for the body text from a canned list, and append the sender name to complete the salutation.  \\
    iv) Assemble the message: combine the greeting, message and salutation from steps i), ii) and iii) to form the final email message. \\
    v) Assign an email ID to the generated email.
\end{enumerate}

\subsection{Additional text generation results }\label{appendix-text}
Here we include some additional results from Section \ref{text-gen-results}.\\
\subsubsection{Subject generation}\label{appendix-text1}
Another example employee was ID 5, who appears to be a person that works in sales or perhaps the legal department. Figure \ref{sales-guy} shows their keywords and generated subjects.
\begin{figure}
\begin{center}
\begin{tabular}{|l|}
\hline
Sale of Napoleonville land.                                  \\ \hline
Price and Interest Rates, as seen on the MarketWatch.        \\ \hline
City of Mesa Update.                                         \\ \hline
Contract or Training.                                        \\ \hline
Sale of the Hines Hines/Nerdwood.                            \\ \hline
Agreement with PG\&E.                                        \\ \hline
Deal Request - M5B17.1.                                      \\ \hline
Contract for:  the Office of the Chairman, and for the  oom. \\ \hline
Agreement with Drexel Energy.                                \\ \hline
Contract Payment Status Report.                              \\ \hline
Deal Correction Notice - Week of Oct 25.                     \\ \hline
Agreement with EPMI-ECI.                                     \\ \hline
\end{tabular}
\end{center}
\caption{Keywords and generated subjects for ID 5.}\label{sales-guy}
\end{figure}

\subsubsection{PPLM content generation}
Figure \ref{PPLM} shows the difference in PPLM generated content when the step size is changed. While the smaller step size results in content of reasonable quality, the larger step size leads to degeneration.

\begin{figure}
\begin{center}
\begin{tabular}{|l|p{0.7\linewidth}|}
\hline
\multicolumn{2}{|l|}{Prompt: Meeting today at 10:30 am PDT. Hi Linda,}\\ \hline
\multicolumn{2}{|l|}{\scell{Subject: Meeting today at 10:30 am PDT.}{From: Sandra Lynch To: Linda Perez}}\\ \hline
\multirow[b]{2}{*}[-6em]{Step size = 0.3} & {Hi Linda,}\\
& {We are meeting at the 10th-16th of November in San Francisco, CA to discuss the development on the East Coast of ENE. As such, the first meeting will be with Linda Adrienne Young and myself to discuss all the features of our plan, what we expect to see, the challenges, and what we can be grateful for in the next 2 weeks. Linda and his team will be meeting with the ENE developers and we will then move on to the 3-4 Meetings after our last meeting.
I have attached a list of what we expect to see, and some of the features and opportunities in the 3-4 Meeting. Sandra.} \\ \hline
\multirow[b]{3}{*}[-3em]{Step size = 0.5} & {Hi Linda,}\\
& {You have a number of outstanding projects that we are working on. I would encourage all of you to contact us today to take a look at this project if you have questions. Meeting Attendees   {[}IMAGE{]} {[}IMAG= E{]}  {[}IMAGE{]}    {[}IMAGE{]} {[}IMAGE{]} {[}IMAGE{]}    {[}IMAGE{]}{[}IMAGE{]}     {[}IMAGE{]} {[}IMAGE{]}  Good Meeting Meeting} \\
& {Regards, Sandra.} \\
\hline
\end{tabular}\caption{email generation using PPLM.}\label{PPLM}
\end{center}
\end{figure}
\subsubsection{Example thread from Enron dataset}\label{appendix-enron-thread}
Figure \ref{tab:enron} shows an example email thread between 3 people which was taken from the Enron dataset. Figure \ref{coherence-matrix} shows the coherence matrix for this thread, which computes the similarity score between all email pairs in the thread from Figure \ref{tab:enron} (instead of just consecutive email pairs as in Section \ref{text-gen-results}). We see that Email 3, which is a forward-type email saying `fyi. Comment?' has a very low similarity scores (below 0.1) with all emails in the thread, whilst the other pairs have similarity scores between 0.4 and 0.51. 

\newcommand{\mystrut}{\rule[-8pt]{0pt}{15pt}\vspace{2.2em}}
\begin{figure}
\begin{center}
\begin{tabular}{|p{0.175\linewidth}|p{0.75\linewidth}|}
\hline
{\textbf{Subject:}} & {\textbf{Escalade}} \\ \hline
Email{\color{white}.}1: New thread \mystrut & \\
  {From: Lauren}\mystrut &\\ 
  {To: Kate}\mystrut &\\ 
  {CC: Neil}\mystrut & \multirow{-4}[tb44]{=}{{Neil and I went to D Taylor and looked at an Escalade. We liked it, however, we want to price it with a couple of extra features. I like the extra wood package, and Neil likes it with the gold trim package. When we asked for a price, it added around \$1700, which I thought was ridiculous. They offered to discount it by \$200, which I again thought was ridiculous. When we priced these options at McGinniss, the price was less (\$600 for the wood package instead of \$995, for instance), and they offered to discount the options by a third.  So, we can get the deal done if they will get reasonable over these options.  The sales guy  (Nathaniel) was going to work on it. The fellow you had told us to ask for wasn't in, by the way. I will be out of town on Monday, but Neil will be in, and will have my Lexus.}} \\
\hline
\renewcommand{\mystrut}{\rule[-4pt]{0pt}{5pt}\vspace{0.5em}}
Email{\color{white}.}2: Reply \mystrut & \\
  {From: Kate}\bigstrut\vspace{0.5em} &\\ 
  {To: Lauren}\bigstrut & \multirow{-3}[tb13]{=}{The dealership will not come down on the price for the gold and wood trim. I have an aftermarket company that can do both for \$550.00. Here is a 36 month quote for the Cadillac including the gold and wood. Please review and let me know if you have any questions.} \\
\hline
Email{\color{white}.}3: Fwd \vspace{0.2em}& \\
  {From: Lauren} \vspace{0.2em}&\\ 
  {To: Neil} \vspace{0.2em}& \multirow{-3}[tb12]{=}{fyi.  Comment?}\\
\hline
\renewcommand{\mystrut}{\rule[-8pt]{0pt}{15pt}\vspace{3.3em}}
Email{\color{white}.}4: Reply \mystrut & \\
  {From: Neil}\mystrut\vspace{3em} &\\ 
  {To: Kate}\bigstrut & \multirow{-3}[tb45]{=}{We need to verify this is a 36 mos lease because  the term value says 48 mos. I had said we wanted to see a rate at 36 mos with 10K down. I assume she used the 10K as total out of pocket expenses at time of delivery and that's why the cost reduction is 7500. The mo payment is about the same as the Lexus. As for the two additions, the aftermarket folks do good work so that should not be a concern. Shoot, some dealerships may even use the same folks to do the work for them and then mark it up. Bottom line, it's your call. If you choose to walk away, I'm ok with it. If you do, what do you want to do going forward. Look at something else? Go back to McGinnis?} \\ \hline
\end{tabular}
\caption{An example email thread between 3 people, taken from the Enron dataset.}\label{tab:enron}
\end{center}
\end{figure}

\begin{figure}
\begin{flushright}
\includegraphics[width=7cm]{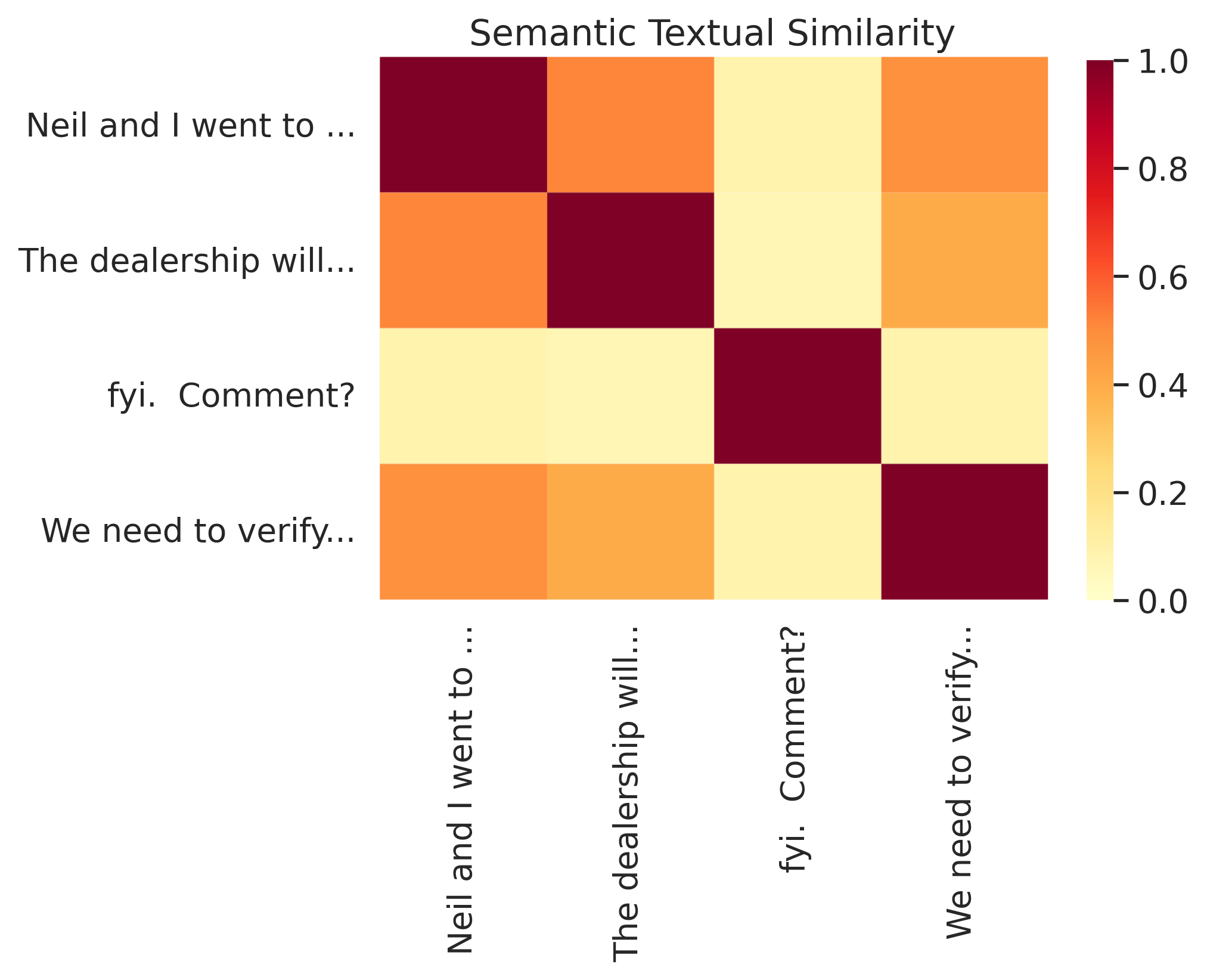}
\caption{Coherence matrix between email pairs in the email thread from Figure \ref{tab:enron}}\label{coherence-matrix}
\end{flushright}
\end{figure}

\section{Discussion continued}
\subsection{Applications of our framework}\label{discussion-applications}
\textbf{Applying pre-trained TPPs in cyber deception for flexible and controllable simulation of social networks} When using a trained TPP model to simulate a social network, the generated events will mimic the underlying network dynamics of the dataset the model was trained on. The lack of publicly available datasets for private communications like email, SMS and instant messaging presents a barrier to developing a diverse range of social network event generation models. Pre-trained language models like GPT have proven to be flexible and versatile tools for language generation, that can be fine-tuned and used for various language applications like sentiment analysis, document categorization and entity extraction\footnote{https://beta.openai.com/docs/guides/fine-tuning}, and even for use with different data types (ImageGPT). Controlling the properties of generated images is another popular research topic \cite{kingma2018glow, karras2019style, zhu2021and, shoshan2021gan}. With this in mind, we ask if a pre-trained TPP model has the flexibility to allow the user deploying the model to control aspects of the simulation such as the number of people in the network, the habits of certain people in the network and the overall tempo and periodicity of communications. This is a problem we have not seen in the literature before. 

The classical parametric TPP framework lends itself naturally to this problem, as adjusting parameters in an individual’s intensity function leads to prescribed and intuitive changes in the temporal dynamics of their interactions. Furthermore, the addition or removal of users can be achieved by increasing or decreasing the size of the parameter vectors. For neural network TPP architectures, this is not as straightforward to achieve. There are tens of thousands of parameters in the model and no obvious interpretation of how each individual one affects the network dynamics. An opportunity for future research is to investigate methods for disengangling the latent space of neural TPP models for the puprose of controlled network simulation.

\textbf{Application of this work to cyber range simulation}
Cyber range environments are another rapidly growing application domain for the use of realistic simulations. Cyber ranges have a host of important applications, including security testing and research, security education and capability development\cite{ECSO2020}, all of which rely on content generation to make each environment distinct and convincing. An example within this evolving field is that of Autonomous Cyber Operations (ACO). As of today, there are a number of network environment frameworks, much like cyber ranges, for the training of autonomous cyberdefence and pentesting agents~\cite{standen2021cyborg,molina2021network,microsoft2021cyberbattlesim,schwartz2019autonomous}. Making ACO environments as realistic as possible is vital to reducing the 'reality gap' between an agent's performance on the simulation environments they're trained on compared to a real network environment, since simulated environments abstract away information that may be critical to an agent’s effectiveness~\cite{standen2021cyborg}. 

\subsection{Communication Generation}\label{discussion-communication}
\textbf{Network evolution over time}
Over time it is natural for an organisation's members to change, and one may wish to reflect this in their email simulation model. In particular, if a company is growing in size it would be desirable to add users to the network simulation. If the original trained TPP model simulates a network of size $n$, creating a larger network of size $o > n$ requires the addition of new marks to the mark embedding space, as well as an increase in the overall tempo of network generation, to account for the larger network size. It would be possible to extend the code in future work to enable this to be achieved by fine-tuning the existing pre-trained model on training data of the target network size. Specifically, this would involve loading the model and removing the last layer of the sender and recipient modules, then changing both those layers to the correct number of neurons for the updated classifier sizes. Fine-tuning with the new data then amounts to training the weights of these classification final layers from scratch. For this reason we believe it is more practical to periodically train the entire model again from scratch. For the alternative case of removing a user in an existing trained TPP, this can be achieved by rejecting any sampled events that include them, or re-training the model again from scratch. 

\textbf{PPLM for multi-person conversation generation}
We also experimented with using PPLM, though did not find it to outperform the standard GPT-2 model. Degeneration is a recognised issue with the PPLM, and discussed in the paper\cite{dathathri2019plug}, and we found it to be particularly noticeable when the step size is set too high or top-k is set too low. Figure \ref{PPLM} in Appendix \ref{appendix-text} shows an example of the difference observed in output when the same prompt is used but different step size. 

An advantage of using PPLM over generating emails without is that if there was a key piece of information that an intruder would be interested in, relevant words could be added to the BOW. So, if the intruder conducted a search for some of these relevant words the fake emails would be of interest to them. PPLM also gives some control as to the topic of the emails, which is important for the goal of enticing intruders. On the other hand, if the intruder were to do a quick scan of an email to determine its legitimacy, the standard GPT-2 model without PPLM tends to sound more natural. Despite the overall improvement of degeneration of PPLM with the tweaking of parameters, there were still some cases where this was an issue. It can particularly be seen when the model produces dates and times – it often degenerates to a list of dates and times regardless of the previous context. 

Another part of our investigations were how to select the "best" sample when generating email bodies. We typically asked the model to generate a number of outputs, and then selected one of them to use. A natural metric for this task is perplexity, with the default choice being the sample with the lowest perplexity. When using the PPLM model however, we found that generated content with degeneration was scored disproportionately well, above arguably more coherent output. This resulted in many simulations that contained not very coherent emails. Generation was also found to be much slower using PPLM than standard GPT-2.

\end{document}